\documentclass[aps,prl,twocolumn,superscriptaddress,10pt]{revtex4-1}
\usepackage[utf8]{inputenc}
\usepackage{amsmath}
\usepackage{amssymb}
\usepackage{graphicx}
\usepackage{pdfpages}
\usepackage{multido}
\makeatletter
\AtBeginDocument{\let\LS@rot\@undefined}
\makeatother

\usepackage{color}
\usepackage{verbatim}
\usepackage{booktabs}
\usepackage{float} 
\usepackage{listings}
\usepackage{alltt}
\usepackage{subfigure}

\usepackage{silence}
\usepackage{hyperref}
\usepackage{cleveref}

\usepackage[normalem]{ulem}

\crefname{equation}{Eq.}{Eqs.}
\Crefname{equation}{Equation}{Equations}
\crefname{figure}{Fig.}{Figs.}
\Crefname{figure}{Figure}{Figures}
\crefname{section}{Sect.}{Sects.}
\Crefname{section}{Section}{Sections}

\newcommand{\ket}[1]{| #1 \rangle}

\newcommand{\ketbra}[1]{| #1 \rangle\langle #1 |}

\newcommand{\expo}[1]{\text{e}^{ #1 }}
\newcommand{\appropto}{\mathrel{\vcenter{
  \offinterlineskip\halign{\hfil$##$\cr
    \propto\cr\noalign{\kern2pt}\sim\cr\noalign{\kern-2pt}}}}}

\newcommand{\ha}{\hat{a}}
\newcommand{\had}{\hat{a}^\dagger}

\newcommand{\hx}{\hat{x}}
\newcommand{\hp}{\hat{p}}

\newcommand{\hbd}{\hat{b}^\dagger}
\newcommand{\hb}{\hat{b}}
\newcommand{\hU}{\hat{U}}
\newcommand{\hH}{\hat{H}}
\newcommand{\hd}{\hat{d}}
\newcommand{\hdd}{\hat{d}^\dagger}

\begin{document}

\title{Stabilization of Finite-Energy Gottesman-Kitaev-Preskill States}
\author{Baptiste Royer}
\affiliation{Department of Physics, Yale University, New Haven, Connecticut 06520, USA}

\author{Shraddha Singh}
\affiliation{Department of Applied Physics, Yale University, New Haven, Connecticut 06520, USA}

\author{S.M. Girvin}
\affiliation{Department of Physics, Yale University, New Haven, Connecticut 06520, USA}


\begin{abstract}
We introduce a new approach to Gottesman-Kitaev-Preskill (GKP) states that treats their finite-energy version in an exact manner. Based on this analysis, we develop new qubit-oscillator circuits that autonomously stabilize a GKP manifold, correcting errors without relying on qubit measurements. Finally, we show numerically that logical information encoded in GKP states is very robust against typical oscillator noise sources when stabilized by these new circuits.
\end{abstract}

\maketitle

In order to build a quantum computer, fragile quantum information must be protected against environmental noise. Quantum error correcting (QEC) codes achieve this through a redundant encoding such that small errors can be detected and corrected before they corrupt the information. The traditional approach to QEC consists in encoding a logical qubit in a large number of physical two-level systems. 
In contrast, bosonic codes take advantage of the large Hilbert space of a single high quality factor harmonic oscillator to encode logical information, providing a hardware-efficient approach to QEC. 
Arguably, the bosonic encodings most pursued experimentally are rotation symmetric codes~\cite{Grimsmo20a} such as the binomial~\cite{Michael16a,Hu19a} and cat codes~\cite{Mirrahimi14a,Ofek16a,Lescanne20a,Grimm20a}. Here, we focus instead on a code developed in the seminal work of Gottesman, Kitaev and Preskill (GKP)~\cite{Gottesman01a}. When subjected to an amplitude damping channel, the GKP encoding was shown to perform favorably compared to the other bosonic codes~\cite{Albert18a,Noh19b}.
However, while simulations indicate that GKP states constitute an attractive option for a robust encoding of logical quantum information, their practical realization remains a challenge. Only recently have individual GKP code words been prepared in the motional mode of a trapped ion~\cite{Fluhmann19a} and the GKP QEC code space been stabilized in a microwave cavity~\cite{Campagne-Ibarcq20a}. In particular, no experiment has shown all the ingredients required for the original fault-tolerant error-correction strategy of Ref.~\cite{Gottesman01a}.
One under appreciated challenge with the stabilization of finite-energy GKP states is that error-correction strategies tailored on the ideal code do not properly take into account the energy injected into the oscillator during error-correction steps, and most theoretical work has instead been focused on the preparation of GKP states~\cite{Travaglione02a,Pirandola04a,Vasconcelos10a,Terhal16a,Motes17a,Shi19a,Hastrup19a,Weigand20a}.

In this Letter, we introduce a new approach to finite-energy GKP states. Concisely, we propose an \emph{exact} approach to finite-energy GKP states instead of considering them as approximate versions of the ideal, unphysical GKP states. From this new approach, we propose new qubit-oscillator circuits that stabilize finite-energy GKP manifolds, allowing the correction of errors in an autonomous fashion.  
Finally, we show numerically that, using these circuits, the lifetime of logical information can be drastically extended. While qubit circuits for the stabilization of the GKP code space were introduced and experimentally realized in Ref.~\cite{Campagne-Ibarcq20a}, we show how these circuits can be put in a broader context and, moreover, we introduce new circuits that yield significantly improved results. 

Since harmonic oscillators are ubiquitous in physical platforms, our approach could be deployed in a variety of systems. The individual ingredients of our protocol have already been demonstrated in a microwave cavity~\cite{Campagne-Ibarcq20a} and in the mechanical motion of a trapped ion~\cite{Fluhmann19a}, and could also be implemented in co-planar waveguide resonators~\cite{Wallraff04a}, bulk acoustic resonators~\cite{Chu17a} or photonic cavities~\cite{Kimble98a,Mabuchi02a}. The circuits we introduce in this Letter could additionally be used to encode information in photon modes used for long-distance quantum communications~\cite{Noh19b}, or to encode oscillator states into oscillators~\cite{Noh20a}.
Finally, beyond the development of quantum computers and quantum communications, GKP states are also a powerful resource state for quantum-enhanced sensing, allowing to probe small forces in arbitrary directions~\cite{Duivenvoorden17a,Zhuang20a}.

\textit{Ideal GKP code}\textemdash
\begin{figure}[t]
    \centering
    \includegraphics[scale=0.65]{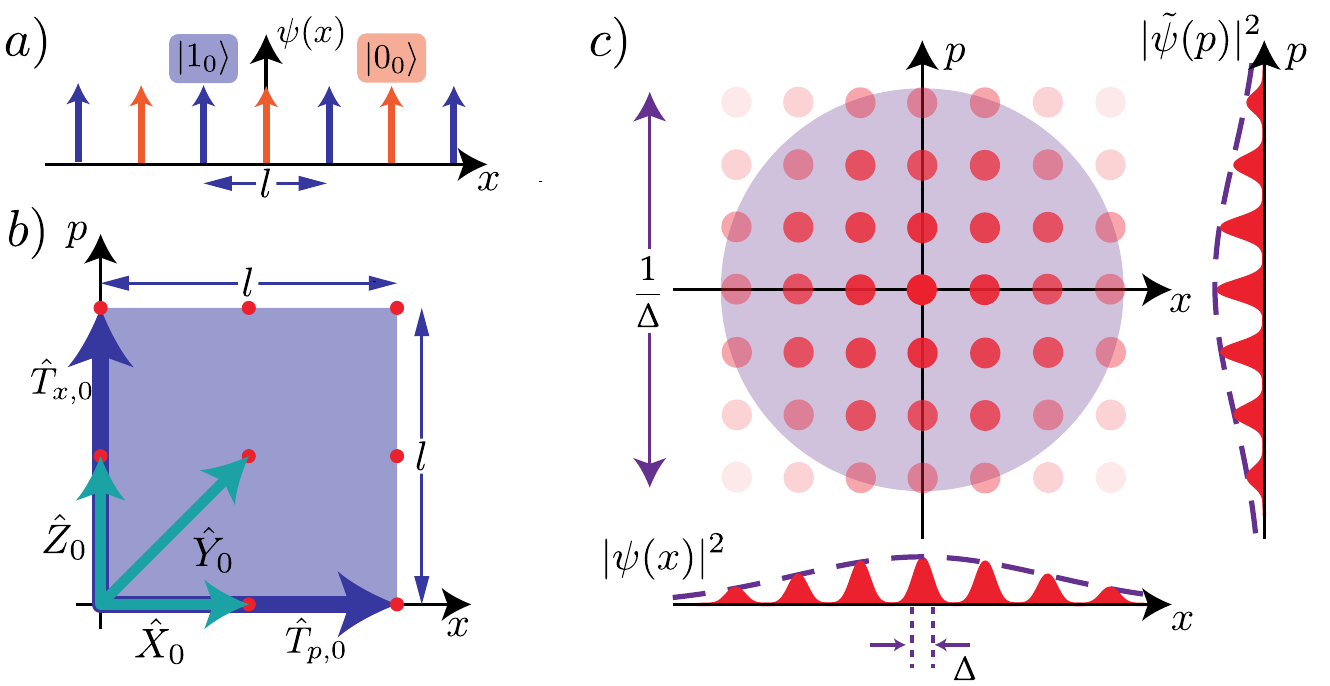}
    \caption{a) Ideal GKP code words consists in superposition of position eigenstates separated by $l$. b) Phase space representation of the ideal stabilizers (blue) and logical Pauli operators (teal). c) Phase space representation of the finite-energy GKP code projector, $\hat P_\mathcal C$. The code lives on grid points $(x,p) = l/2\times (n_1,n_2)$, $n_1,n_2 \in \mathbb Z$ (red), weighted by a Gaussian envelope of width $\sim 1/\Delta$ (purple). The marginals on the bottom and right illustrate that peaks of the ideal code are replaced by Gaussians of width $\sim \Delta$.}
    \label{fig:GKPconcept}
\end{figure}
Before discussing finite-energy GKP states, we start by recalling the core concepts of the ideal GKP error-correction code~\cite{Gottesman01a}. We consider a harmonic oscillator with annihilation and creation operators $\ha$ and $\had$ obeying the commutation relation $[\ha,\had] = 1$ and we denote the quadrature coordinates $\hx = (\ha + \had)/\sqrt 2$ and $\hp = -i(\ha - \had)/\sqrt 2$. The idea of the GKP code is to encode quantum information in translation-invariant grid states and, for simplicity, we consider a square lattice encoding with lattice constant $l = 2\sqrt{\pi}$. We refer the reader to the Supplementary Material~\cite{SM} for a description of GKP states in general lattices.
The ideal code words are defined as the +1 eigenstates of the stabilizers $\hat T_{x,0} = \expo{i l \hat x}$ and $\hat T_{p,0} = \expo{-i l \hat p}$ which,
as illustrated in \cref{fig:GKPconcept}b), translate the wave function by $l$ in the $\hp$ and $\hx$ quadrature, respectively. 
In the $\hx$ quadrature eigenbasis and as illustrated in \cref{fig:GKPconcept}a), the ideal code words $\mu \in \{0,1\}$ are given by Dirac combs, $\ket{\mu_0} \propto \sum_{j\in \mathbb Z} \ket{(j + \mu/2) l}_x$. 
Logical Pauli operators, illustrated in \cref{fig:GKPconcept}b), are given by the translation operators $\hat X_{0} = \expo{-i \frac{l}{2} \hat p}$ and $\hat Z_{0} = \expo{ i \frac{l}{2} \hat x}$.
To describe the GKP code, it is convenient to use modular quadratures (the so-called Zak basis~\cite{Zak67a}) $\hx_{[m]} = \hx \text{ mod } m$ and $\hp_{[m]} = \hp \text{ mod } m$, here taken in a symmetric interval around 0: $x_{[m]},p_{[m]} \in (-m/2,m/2]$~\cite{Pantaleoni20a}. In particular, the stabilizer condition can be equivalently expressed as $\hat T_{x,0}\ket{\psi} = \ket{\psi} \Leftrightarrow \hx_{[l/2]} \ket{\psi} = 0$. 

Consider a state evolving under a noise channel consisting in random displacements. To correct such errors, the standard approach is to measure $\hx_{[l/2]}$ and $\hp_{[l/2]}$, say with results $u$ and $v$, respectively. After these measurements, the state is in an eigenstate of the stabilizers $\hat T_{x,0},\hat T_{p,0}$ with eigenvalues $\expo{ilu},\expo{-ilv}$, respectively, and errors are corrected by applying a displacement $\hat D[-(u + i v)/\sqrt{2}]$, with $\hat D(\alpha) = \exp\{\alpha \had - \alpha^* \ha\}$~\cite{C.W.-Gardiner00a,Gottesman01a}. As long as errors are small, $u,v < l/4$, the logical information is perfectly recovered: GKP states are therefore robust to any error channel that corresponds to a superposition or mixture of small displacements~\cite{Gottesman01a,Glancy06a}.

Importantly, during the procedure above, the state is projected onto the infinitely squeezed eigenstates of the stabilizers which contain an infinite amount of energy. While these idealized measurement of the modulo quadratures are not possible since physical measurements have finite precision, the back action of any realistic $\hx_{[l/2]}$ or $\hp_{[l/2]}$ measurement results in increased squeezing and a corresponding accumulation of energy in the oscillator. 
For noise channels independent of the excitation number such as classical phase space diffusion channels, this energy increase is not detrimental.
However, typical oscillator error channels such as amplitude damping do scale with energy and, furthermore, harmonic oscillators usually inherit some form of non-linearity due to their coupling with a control element required for full quantum control of the oscillator mode. For example, microwave cavities inherit a self-Kerr non-linearity from their coupling with a superconducting qubit, and state-dependent forces on the mechanical motion of trapped ions can depend on the phonon number due to the breakdown of the Lamb-Dicke approximation~\cite{Wineland98a}. It is thus crucial to design error-correction strategies that control squeezing in GKP states in order to limit the effect of noise channels that scale unfavorably with excitation number.

\textit{Finite-energy GKP states}\textemdash
In order to describe physical GKP states, we use an envelope operator $\hat E_\Delta = \exp\{-\Delta^2 \had \ha\}$~\cite{Menicucci14a},
\begin{equation}
\label{eq:finiteEnergyStates}
\ket{\mu_\Delta} = \mathcal N_\Delta \hat E_\Delta \ket{\mu_0},
\end{equation}
where $\Delta$ parametrizes the size of the GKP state and $\mathcal N_\Delta$ is a normalization factor~\cite{SM}. The phase space representation of the finite-energy code space projector, $\hat P_\mathcal C = \ketbra{0_\Delta} + \ketbra{1_\Delta}$, is schematically illustrated in \cref{fig:GKPconcept}c). When $\Delta \ll 1$, the finite-energy GKP states differ from their ideal counterpart only by (a coherent superposition of) correctable errors~\cite{SM}. Consequently, the states defined in \cref{eq:finiteEnergyStates} can be treated as good approximations of the ideal GKP states, and as such retain their QEC properties. However, this approximate approach does not take into account changes in GKP size due to error-correction steps.

To remedy this situation, we define new finite-energy stabilizers through the similarity transformation induced by the envelope operator~\cite{SM}
\begin{subequations}\label{eq:finiteEnergyStabilizers}
\begin{align}
\hat T_{x,\Delta} &= \hat E_\Delta \hat T_{x,0} \hat E_\Delta^{-1} = \expo{i l (c_\Delta \hx + i \hp s_\Delta)},\\
\hat T_{p,\Delta} &= \hat E_\Delta \hat T_{p,0} \hat E_\Delta^{-1} = \expo{-i l (c_\Delta \hp - i \hx s_\Delta)},
\end{align}
\end{subequations}
where we have defined $c_\Delta = \cosh\Delta^2$, $s_\Delta = \sinh\Delta^2$ and $t_\Delta = \tanh\Delta^2$.
Crucially, the finite-energy states are exact +1 eigenstates of these new stabilizers, $\hat T_{x,\Delta}\ket{\mu_\Delta} = \hat T_{p,\Delta}\ket{\mu_\Delta} = \ket{\mu_\Delta}$. We also define finite-energy Pauli operators in a similar way, $\hat X_{\Delta} = \hat E_\Delta \hat X_{0} \hat E_\Delta^{-1}$ and $\hat Z_{\Delta} = \hat E_\Delta \hat Z_{0} \hat E_\Delta^{-1}$. Although finite-energy operators are neither unitary nor Hermitian, they obey the same commutation relations as their ideal counterpart. Moreover, this finite-energy approach can be generalized to different envelope shapes~\cite{SM}.

From the squeezed annihilation operator $\ha_{x,\Delta} = \hat S^\dag(\ln \sqrt{t_\Delta}) \ha \hat S(\ln \sqrt{t_\Delta})$, we write $\hat T_{x,\Delta} = \exp\{il \sqrt{2s_\Delta c_\Delta} \ha_{x,\Delta} \}$, with $\hat S(\xi) = \exp\{\xi^*/2\ha\ha - \xi/2\had\had)\}$ the standard squeezing transformation~\cite{C.W.-Gardiner00a}. In other words $\hat T_{x,\Delta}$ is a function of $\ha_{x,\Delta}$, and consequently shares the same eigenstates which are squeezed coherent states. This relation highlights the fact that the finite-energy GKP states defined in \cref{eq:finiteEnergyStates} are equivalent to superpositions of finitely squeezed states~\cite{Matsuura19a}. Moreover, we can also write
\begin{equation}\label{eq:darkStateFiniteEnergy}
\hat T_{x,\Delta}\ket{\psi} = \ket{\psi} \Leftrightarrow \hd_{x,\Delta}\ket{\psi} = 0,
\end{equation}
where $\hd_{x,\Delta} = -i/(l\sqrt{2s_\Delta c_\Delta})\ln \hat T_{x,\Delta} = (\hx_{[l/2c_\Delta]}/\sqrt{t_\Delta} + i \hp\, \sqrt{t_\Delta})/\sqrt 2$. The $\hd_{x,\Delta}$ operator differs from $\ha_{x,\Delta}$ in that the $\hx$ quadrature is replaced by its modular counterpart due to the multivalued nature of the complex logarithm. 

\Cref{eq:darkStateFiniteEnergy} suggests an improved strategy for error-correction: instead of measuring $\hx_{[l/2]}$, which would increase the amount of energy in the GKP state, one option is to engineer an oscillator-bath interaction such that the oscillator naturally relaxes to the dark states $\hd_{x,\Delta}\ket{\psi} = 0$. Bath engineering approaches~\cite{Albert16a} have been successfully used for the stabilization of single squeezed states~\cite{Kienzler15a} and in the QEC context for the stabilization of cat states~\cite{Mirrahimi14a,Touzard18a,Lescanne20a} and binomial code words~\cite{Gertler20a}.
Here, we propose that finite-energy GKP states can be stabilized by engineering an effective Markovian oscillator-bath interaction 
\begin{equation}\label{eq:coolingHamiltonian}
\hH_{x,\Delta}(t) = \sqrt{\Gamma_x} [\hd_{x,\Delta} \hbd_\tau + \hdd_{x,\Delta} \hb_\tau],
\end{equation}
where the bath operators $\hb_\tau$ and $\hbd_{\tau'}$ obey $[\hb_\tau,\hbd_{\tau'}]=\delta(\tau-\tau')$ and $\langle \hbd_\tau\hb_{\tau'}\rangle = 0$. The above Hamiltonian describes a process where ``elementary excitations'' of the GKP states, created by $\hdd_{x,\Delta}$, are transferred to a zero temperature bath, autonomously cooling the harmonic oscillator towards the +1 eigenspace of $\hat T_{x,\Delta}$ at a rate $\propto \Gamma_x$. By engineering a second bath with an analogous interaction for the $\hat T_{p,\Delta}$ stabilizer, the finite-energy GKP manifold is autonomously stabilized.
While this oscillator-bath interaction seems challenging to implement, we show how it can be approximated by using an ancilla qubit with levels denoted $\{\ket{g},\ket{e}\}$.

\begin{figure}[t]
    \centering
    \includegraphics[scale=0.61]{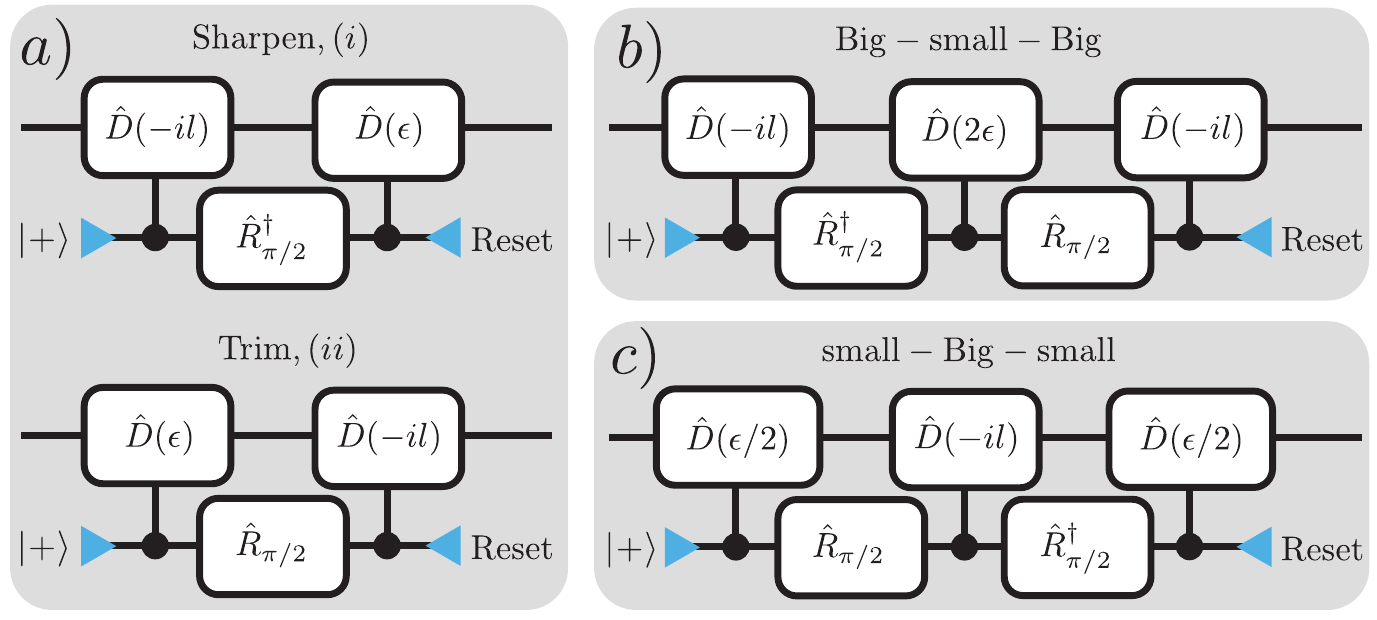}
    \caption{Three protocols that autonomously stabilize the +1 eigenspace of $\hat T_{x,\Delta}$. In all circuits, ancillas (bottom wire) are initialized in $\ket{+} = (\ket{g} + \ket{e})/\sqrt 2$ and reset at the end. Ancilla rotations are given by $\hat R_{\pi/2} = \exp\{-i \hat \sigma_x \pi/4\}$. a) Autonomous version of the Sharpen-Trim circuits introduced in Ref.~\cite{Campagne-Ibarcq20a}. The two steps from ST can be combined into the BsB (b) or sBs (c) protocols.}
    \label{fig:stabilizationCircuit}
\end{figure}

\textit{Qubit Stabilization}\textemdash
Concisely, we discretize the time evolution generated by the Hamiltonian \cref{eq:coolingHamiltonian} and replace the bath by a qubit that is reset between each step, an approach that can also mimic standard models of dissipation~\cite{Brun02a,Bouten09a,Ciccarello17a,Gross18a}.
First, consider a discretization of the bath modes over slices of time $\delta t$ small enough such that the average number of excitations transferred to the bath in one time step is much smaller than 1. As a result, the bath can be replaced by an ensemble of ancilla qubits, $\hb_\tau\rightarrow (\hat \sigma_{x,t} + i \hat \sigma_{y,t})/\sqrt{2\delta t}$, where $\hat \sigma_{\alpha,t}$ are the Pauli matrices of the t$^{\mathrm{th}}$ qubit, with $\alpha\in\{x,y,z\}$ and $t\in \mathbb Z$. Since ancillas are discarded after interacting with the oscillator, we take a single qubit (dropping the $t$ index) that is reset between each step. The desired oscillator subspace is thus stabilized by repeatedly applying the discretized unitary~\cite{SM}
\begin{equation}\label{eq:targetU}
\hU = \exp\left\{-i\sqrt{\frac{\Gamma_x \delta t}{t_\Delta}}\left(\hx_{[l/2c_\Delta]}\hat \sigma_x + \hp \hat \sigma_y t_\Delta\right)\right\}.
\end{equation}
The implementation of \cref{eq:targetU} can be simplified by leveraging the inherent modularity of the qubit.
First, we make a Trotter decomposition to separate the terms with the modular ($\hx_{[l/2c_\Delta]}$) and standard ($\hp$) quadratures. Then, we replace $\hx_{[l/2c_\Delta]} \rightarrow \hx$, choosing $\Gamma_x \delta t$ such that a translation $\hx \rightarrow \hx + l/2c_\Delta$ leads to a trivial qubit operation after the whole interaction is completed.
From different Trotter decompositions of the target unitary \cref{eq:targetU}, we find three protocols:   
\begin{subequations}\label{eq:stabilizationCircuits}
\begin{align}
\text{Sharpen-Trim: }\hU^{(ST)} &= 
\begin{cases}
      \expo{\frac{-i\epsilon \hp \hat \sigma_y}{2}} \expo{\frac{-il c_\Delta \hx\hat \sigma_x}{2}}, & (i)\\
      \expo{\frac{-il c_\Delta \hx\hat \sigma_x}{2}}\expo{\frac{-i\epsilon \hp \hat \sigma_y}{2}}, & (ii)
\end{cases}\\
\text{Big-small-Big: }\hU^{(BsB)} &= \expo{\frac{-il c_\Delta \hx\hat \sigma_x}{2}}\expo{-i\epsilon \hp \hat \sigma_y}\expo{\frac{-il c_\Delta \hx\hat \sigma_x}{2}},\\
\text{small-Big-small: }\hU^{(sBs)} &= \expo{\frac{-i\epsilon \hp \hat \sigma_y}{4}}\expo{\frac{-il c_\Delta\hx\hat \sigma_x}{2}}\expo{\frac{-i\epsilon \hp \hat \sigma_y}{4}},
\end{align}
\end{subequations}
where we have defined $\epsilon = s_\Delta l \approx \Delta^2 l$. \Cref{eq:stabilizationCircuits}a is obtained through first order decompositions, $\expo{\delta(A+B)} \approx \expo{\delta A}\expo{\delta B} + \mathcal O(\delta^2)$, while b) and c) result from second order Trotter decompositions that approximate better \cref{eq:targetU}, $\expo{\delta(A+B)} \approx \expo{\delta A/2}\expo{\delta B}\expo{\delta A/2} + \mathcal O(\delta^3)$. The ST protocol requires 2 steps, while the BsB and sBs protocols require only 1 step.
Although the protocols in \cref{eq:stabilizationCircuits} could be implemented directly, we choose to write them in terms of a controlled displacement operation, $C\hat D(\beta) = \exp\{(\beta \had - \beta^*\ha) \hat \sigma_z /2\sqrt2\}$, whose
effect is to displace the oscillator $\hx$ ($\hp$) quadrature in a symmetric way by $\pm$Re$[\beta/2]$ ($\pm$Im$[\beta/2]$)  conditioned on the ancilla being in $\ket{g}$ or $\ket{e}$. This operation has been demonstrated in trapped ions~\cite{Haljan05a,Fluhmann19a} and more generally can be implemented with a two-level system coupled dispersively to the oscillator mode~\cite{Campagne-Ibarcq20a}.
Inserting qubit rotations in \cref{eq:stabilizationCircuits}, we obtain the circuits illustrated in \cref{fig:stabilizationCircuit}~\cite{SM}. The circuits for the $\hat T_{p,\Delta}$ stabilizer are obtained by mapping all controlled displacements $C\hat D(\beta) \rightarrow C\hat D(i\beta)$.

A few comments are in order here. First, we note that the ST protocol is an autonomous version of the Sharpen-Trim protocol introduced in Ref.~\cite{Campagne-Ibarcq20a}. There, the values of $\epsilon$ in circuits ($i$) and ($ii$) were optimized independently and an optimal working point was found at $\epsilon_{(i)} = \epsilon_{(ii)}$, in agreement with \cref{eq:stabilizationCircuits}a. 
Moreover, in the ST protocol, the size of the GKP expands during the sharpening step ($i$) and then contracts during the trimming step ($ii$). In contrast, GKP states keep a constant size for both BsB and sBs protocols.
Second, we remark that the continuous limit $\delta t\rightarrow 0$ does not exist since we fixed $\Gamma_x \delta t$.
The only free parameter of \cref{eq:stabilizationCircuits}, $\epsilon$, sets the size of the stabilized GKP manifold, $\Delta$. In practice, $\delta t$ is fixed by the repetition rate of the stabilization circuits and, for a given experimental set-up, the effective cooling rate decreases with increasing GKP size, $\Gamma_x \appropto \Delta^2$. In short, errors are corrected at a slower rate for larger GKP states.
Moreover, since $\epsilon$ is fixed, there are finite $\hp$ displacements that commute with the protocols. Hence, the $\{$ST,BsB,sBs$\}$ protocols simultaneously stabilize two grids: one with the desired lattice constant $l/2$ and one with superlattice constants $\{2\pi/\epsilon,\pi/\epsilon,4\pi/\epsilon\}$, respectively. The effects of the superlattice are minor as long as the GKP state is confined within its central unit cell around $\langle \ha \rangle = 0$. Finally, analyzing the information the final qubit state contains about the oscillator state, the two steps of the ST protocol can be understood as one bit phase-estimation and homodyne measurements, respectively. In contrast, the sBs and BsB protocols perform a simultaneous, one-bit phase estimation of two slightly mismatched quadratures, the superlattice appearing from a Moiré pattern, see the Supplemental Material~\cite{SM}.

\begin{figure}[t]
    \centering
    \includegraphics[scale=0.3]{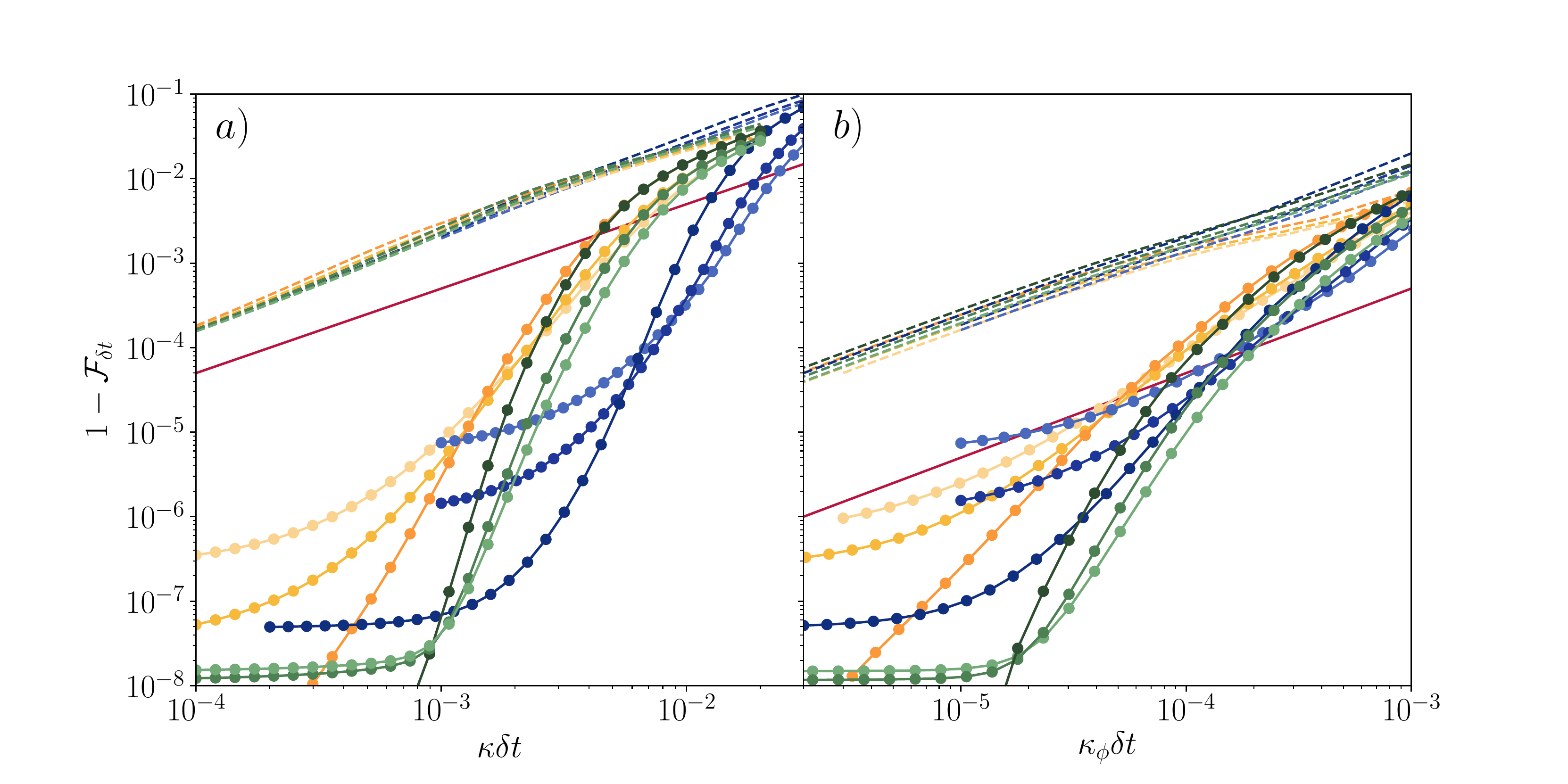}
    \caption{Channel infidelity for a time $\delta t$ in the presence of single excitation loss (a) and oscillator dephasing (b). In both panels, \{orange,blue,green\} curves refer to the \{ST,BsB,sBs\} protocols, while the red curve corresponds to the $\{0,1\}$ Fock encoding. Full (dashed) lines stand for corrected (uncorrected) GKP states. Light to dark colors refer to $\epsilon=\{0.15,0.13,0.1\}$.}
    \label{fig:cavityErrors}
\end{figure}

\textit{Oscillator errors}\textemdash
We now turn to the robustness of the encoded logical information against typical errors.
We first consider perfect and instantaneous stabilization circuits followed by an idle time $\delta t$ where the oscillator evolves under a noise channel.
In order to stabilize the +1 eigenspace of both $\hat T_{x,\Delta}$ and $\hat T_{p,\Delta}$ operators, a full round is completed after $2\delta t$ for the BsB and sBs protocols and $4\delta t$ for the ST protocol. After having stabilized the GKP manifold in an oscillator initialized in vacuum, each of the six Pauli eigenstates are prepared by logical operator measurement~\cite{Campagne-Ibarcq20a,SM}. Then, the decay time of the logical information is computed for each eigenstate, from which the channel fidelity compared to the identity channel is extracted~\cite{SM}. \Cref{fig:cavityErrors} shows the channel infidelity when the oscillator is subjected a) to single-excitation loss at a rate $\kappa$ or b) to cavity dephasing at a rate $\kappa_\phi$.
The theoretical infidelity for the $\{0,1\}$ Fock encoding, the longest-lived encoding without QEC, is shown in red.
We choose $\epsilon = \{0.1,0.13,0.15\}$ such that the mean number of excitations is $\bar n \approx \{12,14,17\}$, respectively. While smaller values of $\epsilon$ are perfectly valid, larger values of $\epsilon$ lead to a small superlattice size and a poor confinement of the stabilized GKP states, especially for the BsB protocol. 
Dashed lines refer to freely evolving GKP states while the full lines correspond to stabilized states. For uncorrected GKP states, the logical error rates scales the same way as the Fock encoding, with an overhead due to the increased number of excitations. 
In contrast, all stabilization protocols enable an extension of the logical information lifetime beyond what is possible with the Fock encoding, with the sBs and BsB protocols showing a clear improvement over ST.
For all protocols the break-even point with respect to the Fock encoding decreases for increasing GKP size and, consequently, there is an optimal $\Delta$ below which performances degrade for a given error channel. In other words, a larger GKP state can handle more errors, but corrects them at a slower rate ($\Gamma \propto \Delta^2$).
For the BsB protocol, the infidelity at small error rates saturates due to ``tunneling'' between sites of the $\pi/\epsilon$ superlattice, which also results in a slow increase in excitation number~\cite{SM}.

\begin{figure}[t]
    \centering
    \includegraphics[scale=0.3]{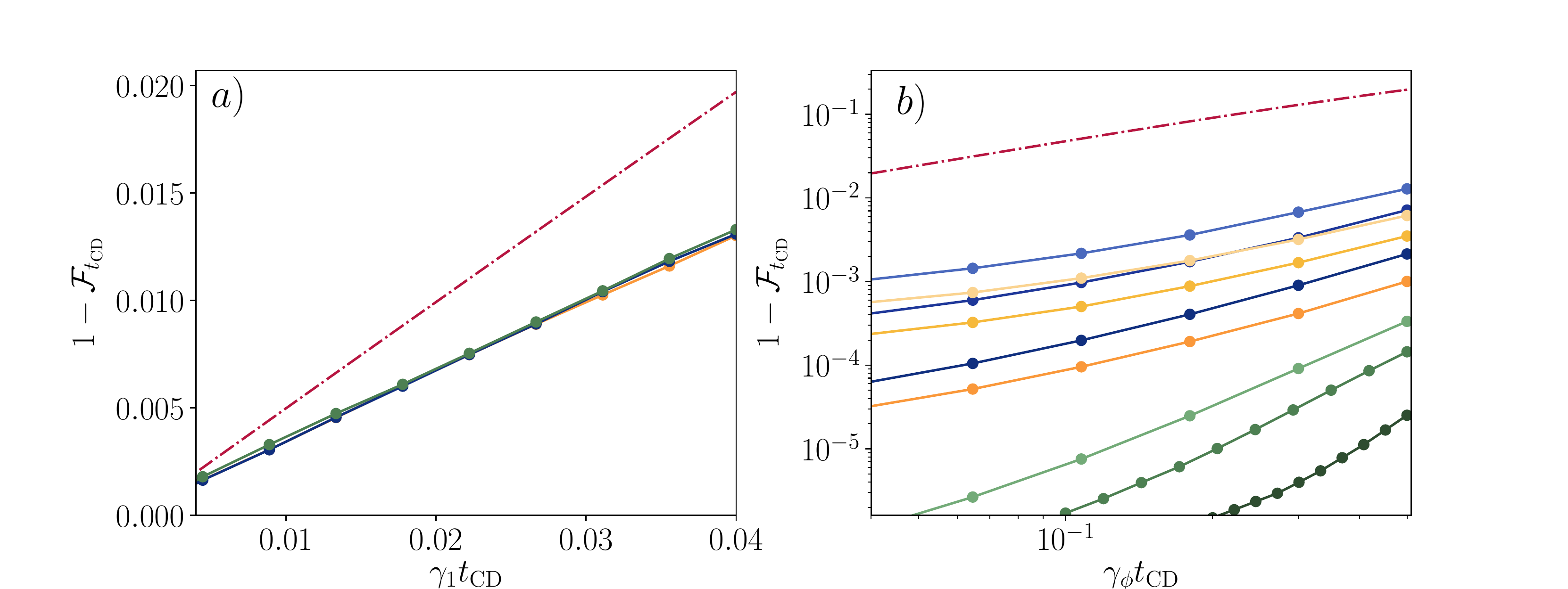}
    \caption{Channel infidelity for a time $t_\mathrm{CD}$ in the presence of ancilla decay (a) and ancilla dephasing (b). The color scheme is the same as in \cref{fig:cavityErrors}, with the dotted-dashed red line referring to the infidelity of the bare ancilla.}
    \label{fig:ancillaErrors}
\end{figure}

\textit{Ancilla errors}\textemdash
In order to study the effect of ancilla errors, we now consider a perfect oscillator and a finite time for stabilization circuits such that a single large controlled displacement lasts for $t_\mathrm{CD}$. \Cref{fig:ancillaErrors} shows the channel infidelity when the ancilla is subjected to a) decay at a rate $\gamma_1$ and b) dephasing at a rate $\gamma_\phi$.
As shown in \cref{fig:ancillaErrors}a), the infidelity is proportional to $\gamma_1$ since decay events \emph{during} large controlled displacements can lead to large displacement errors. Indeed, each $C\hat D(il)$ operation induces a logical error with probability $p /2$, where $p \approx \gamma_1 \delta t$ is the probability of ancilla decay during the controlled displacement~\cite{SM}. These errors, which do not depend on the size of the stabilized GKP, can be partly mitigated for the BsB and ST protocols by replacing the final large controlled displacement by a qubit measurement followed by classical feedback.
As shown in \cref{fig:ancillaErrors}b), the logical information is robust against ancilla dephasing since phase errors commute with controlled displacements. Phase error are converted into bit flips during qubit rotations, but this leads to displacement errors of $\epsilon$ or $l$, both of which are correctable~\cite{SM}.

\textit{Discussion}\textemdash
While in this Letter we focused on the stabilization of square GKP states, the protocols illustrated in \cref{fig:stabilizationCircuit} are flexible and can be extended to the stabilization other lattices (e.g. hexagonal lattices) or $d$-level GKP qudits~\cite{SM}. They can also be adapted to measure more precisely the Pauli operators for finite-energy GKP states~\cite{Hastrup20a,SM}. 
Finally, the lattice can also be reshaped (e.g.\ from square to hexagonal) without losing the encoded information by adiabatically varying the protocol, see the Supplemental Material for more details~\cite{SM}.

As indicated by \cref{fig:ancillaErrors}a), the protocols in \cref{fig:stabilizationCircuit} are not fault-tolerant with respect to ancilla decay. 
An attractive solution that leverages the robustness of the protocols to dephasing errors would be to use a biased-noise ancilla such as the Kerr cat~\cite{Puri19a} where bit flips are suppressed.
Even when ignoring ancilla errors, our stabilization procedures do not provide a way to increase the logical lifetime arbitrarily.
Rather, we envision concatenating the GKP encoding with some other qubit code such as the surface code~\cite{Kitaev03a,Vuillot19a,Noh20b,SM}. By providing a highly-coherent qubit at the base level of the code, we expect the GKP encoding to provide a significant reduction in hardware overhead required for the implementation of a quantum computer.

\textit{Note added}\textemdash While writing our manuscript, we became aware of similar work on the stabilization of GKP states~\cite{Nguyena} and on the measurement of Pauli operators~\cite{Hastrup20a,Nguyena}.

\begin{acknowledgments}
This work was partly inspired by a comment from Mazyar Mirrahimi who brought to our attention a perturbative version of \cref{eq:finiteEnergyStabilizers}. We also thank Philippe Campagne-Ibarcq and Jonathan Home for valuable discussions.
Part of this work was supported by the Army Research Office under Grant No. W911NF-18-1-0212.
\end{acknowledgments}

\bibliographystyle{apsrev4-1}

%

\multido{\i=1+1}{27}{%
		\clearpage
        

\section*{Supplementary material}

# Supplemental Material for
# "Stabilization of Finite-Energy Gottesman-Kitaev-Preskill States"

Baptiste Royer,[1] Shraddha Singh,[2] and S.M. Girvin[1]

[1] *Departement of Physics, Yale University, New Haven, Connecticut 06520, USA*
[2] *Departement of Applied Physics, Yale University, New Haven, Connecticut 06520, USA*

## CONTENTS

## I. NOTATION

In this document, we will consider a harmonic oscillator with annihilation $\hat{a}$ and creation operator $\hat{a}^\dagger$ that respect the commutation relation $[\hat{a}, \hat{a}^\dagger] = 1$. We denote the excitation number operator $\hat{n} = \hat{a}^\dagger \hat{a}$ and the quadrature coordinates

$$
\begin{aligned}
\hat{x} &= \frac{\hat{a} + \hat{a}^\dagger}{\sqrt{2}}, \\
\hat{p} &= -i \frac{(\hat{a} - \hat{a}^\dagger)}{\sqrt{2}},
\end{aligned}
\tag{S1}
$$

respecting the commutation relation $[\hat{x}, \hat{p}] = i$. With this definition, we have $\hat{a} = (\hat{x} + i\hat{p})/\sqrt{2}$. We will make use of the displacement operator [1]

$$
\hat{D}(\alpha) = \mathrm{e}^{\alpha \hat{a}^\dagger - \alpha^* \hat{a}},
\tag{S2}
$$

with $\alpha \in \mathbb{C}$. We will also make use of the squeezing operator

$$
\hat{S}(\xi) = \mathrm{e}^{\frac{\xi^* \hat{a}\hat{a} - \xi \hat{a}^\dagger \hat{a}^\dagger}{2}},
\tag{S3}
$$

with $\xi = r\mathrm{e}^{i\theta}$, whose effect is given by

$$
\hat{S}^\dagger(\xi) \hat{a} \hat{S}(\xi) = \hat{a} \cosh r - \hat{a}^\dagger \mathrm{e}^{i\theta} \sinh r.
\tag{S4}
$$

## II. GENERAL GKP THEORY

### A. Stabilizers

In the main Letter, we focused on the square, qubit GKP code for simplicity. However, the GKP encoding proposed in Ref. [2] is more general and can be defined on other lattices. We start with defining the quadrature coordinates

$$
\hat{q}_j = \alpha_j \hat{x} + \beta_j \hat{p},
\tag{S5}
$$

with $j \in \{1, 2\}$ and $\alpha_j, \beta_j \in \mathbb{R}$. For convenience, we define also the perpendicular quadrature

$$
\hat{q}_{j,\perp} = \alpha_j \hat{p} - \beta_j \hat{x},
\tag{S6}
$$

such that $[\hat{q}_i, \hat{q}_{j,\perp}] = i l_j^2$ with $l_j = \sqrt{\alpha_j^2 + \beta_j^2}$. We take the ideal code stabilizer group generators to be the displacements generated by the $\hat{q}_j$ coordinates,

$$
\hat{T}_{j,0} = \mathrm{e}^{i\hat{q}_j},
\tag{S7}
$$

with $\hat{T}_{j,0} = \hat{D}[(-\beta_j + i\alpha_j)/\sqrt{2}]$. We use the 0 index to describe the ideal code, a choice that will be justified below when describing the finite-energy version of the code. The commutation of the two stabilizers is given by

$$
\hat{T}_{1,0} \hat{T}_{2,0} = \mathrm{e}^{i\omega_{12}} \hat{T}_{2,0} \hat{T}_{1,0},
\tag{S8}
$$

with $\omega_{12} = \alpha_1\beta_2 - \beta_1\alpha_2$. In order to define valid stabilizers, the two translation operators should commute, which imposes the condition $\omega_{12} = 2\pi d$, with $d \in \mathbb{Z}$. This way, the $\hat{T}_{i,0}$ operators generate the stabilizer group of a $d$-level qudit. The generators of translation, $\hat{q}_j$, can be put into matrix form

$$
\begin{pmatrix} \hat{q}_1 \\ \hat{q}_2 \end{pmatrix} = \begin{pmatrix} \alpha_1 & \beta_1 \\ \alpha_2 & \beta_2 \end{pmatrix} \begin{pmatrix} \hat{x} \\ \hat{p} \end{pmatrix},
$$
$$
\vec{\hat{q}} = L \vec{\hat{x}}.
\tag{S9}
$$

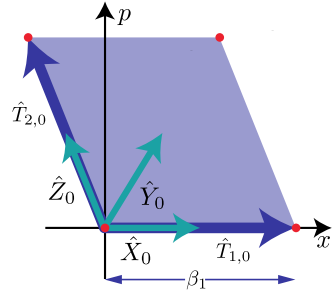

FIG. S1. The elements of the stabilizer group for the GKP encoding are in one-to-one correspondence with the lattice points in red which extend in all directions. The generators of that lattice, $\hat{T}_{1,0}$ and $\hat{T}_{2,0}$, are shown in blue. The condition that the stabilizers commute is equivalent to the condition that the blue shaded space has an area $2\pi d$. The logical Pauli operators are shown in teal.

and the requirement that the stabilizers commute is equivalent to the condition $\det L = \omega_{12} = 2\pi d$. In the following we will focus on the qubit case, $d = 2$. Two popular lattice choices are the square lattice discussed in the main Letter and the hexagonal lattice, which are respectively generated by

$$L_\square = l_\square \begin{pmatrix} 0 & -1 \\ 1 & 0 \end{pmatrix}, \tag{S10a}$$

$$L_\bigcirc = l_\bigcirc \begin{pmatrix} 0 & -1 \\ \sin(\pi/3) & \cos(\pi/3) \end{pmatrix}, \tag{S10b}$$

with $l_\square = 2\sqrt{\pi}$ ($l_\square = l$ in the main Letter) and $l_\bigcirc = \sqrt{8\pi/\sqrt{3}}$.

## B. Logical Pauli operators

The logical Pauli operators are defined from the stabilizers as

$$\hat{X}_0 = \sqrt{\hat{T}_{1,0}} = e^{i\frac{\hat{q}_1}{2}}, \tag{S11a}$$

$$\hat{Z}_0 = \sqrt{\hat{T}_{2,0}} = e^{i\frac{\hat{q}_2}{2}}. \tag{S11b}$$

We verify that the Pauli operators respect the correct commutation relation

$$\begin{aligned} \left[\hat{X}_0, \hat{T}_{2,0}\right] &= \hat{X}_0 \hat{T}_{2,0} - e^{i\frac{\omega_{12}}{2}} \hat{X}_0 \hat{T}_{2,0} = 0, \\ \left[\hat{Z}_0, \hat{T}_{1,0}\right] &= \hat{Z}_0 \hat{T}_{1,0} - e^{i\frac{\omega_{12}}{2}} \hat{Z}_0 \hat{T}_{1,0} = 0, \end{aligned} \tag{S12}$$

with $\omega_{12} = 4\pi$. With the definitions of Eq. (S11), it is immediate that $[\hat{X}_0, \hat{T}_{1,0}] = [\hat{Z}_0, \hat{T}_{2,0}] = 0$. Moreover, we also verify that the Pauli operators anticommute,

$$\begin{aligned} \hat{X}_0 \hat{Z}_0 &= e^{i\frac{\hat{q}_1}{2}} e^{i\frac{\hat{q}_2}{2}}, \\ &= e^{i\frac{\omega_{12}}{4}} e^{i\frac{\hat{q}_2}{2}} e^{i\frac{\hat{q}_1}{2}}, \\ &= -\hat{Z}_0 \hat{X}_0, \end{aligned} \tag{S13}$$

as desired. The final Pauli operator is given by $\hat{Y}_0 = -i\hat{Z}_0\hat{X}_0 = e^{i\frac{\hat{q}_1+\hat{q}_2}{2}}$, as illustrated in Fig. S1.

### C. States

We define the (unnormalized) qubit ideal states in the $\hat{x}$ basis

$$|\mu_0\rangle = \frac{1 + (-1)^\mu \hat{Z}_0}{2} \sum_{j,k \in \mathbb{Z}} \hat{T}_{1,0}^j \hat{T}_{2,0}^k |0\rangle_x. \tag{S14}$$

In the particular case of the square lattice, we obtain (up to an infinite constant)

$$|\mu_0\rangle = \sum_{j \in \mathbb{Z}} |l_\square(j + \mu/2)\rangle_x. \tag{S15}$$

### D. Modular quadratures

In the main text, we used the modular quadrature $\hat{x}_{[m]} = \hat{x} \bmod m$, which can be expressed in the $x$ basis

$$\hat{x}_{[m]} = \sum_{k \in \mathbb{Z}} \int_{-m/2}^{m/2} dx\, x |x + km\rangle\langle x + km|, \tag{S16}$$

or using the Fourier series

$$\hat{x}_{[m]} = \frac{-m}{\pi} \sum_{k=1}^{\infty} \frac{(-1)^k}{k} \sin\left(\frac{2\pi \hat{x} k}{m}\right). \tag{S17}$$

We define the modular $\hat{p}_{[m]}$ quadrature in a similar way. More generally, we can use modular quadratures in the arbitrary directions of $\hat{q}_j$. Rescaling the coordinates $\hat{q}_j = l_j \hat{x}_j$ and $\hat{q}_{j,\perp} = l_j \hat{x}_{j,\perp}$ such that $[\hat{x}_j, \hat{x}_{j,\perp}] = i$, we define $\hat{x}_{j,[m]} = \hat{x}_j \bmod m$,

$$\hat{x}_{j,[m]} = \frac{-m}{\pi} \sum_{k=1}^{\infty} \frac{(-1)^k}{k} \sin\left(\frac{2\pi \hat{x}_j k}{m}\right). \tag{S18}$$

Modular variables and ideal stabilizers are related through the complex logarithm function,

$$\ln \hat{T}_{j,0} = i l_j \hat{x}_{j,[2\pi/l_j]}, \tag{S19}$$

and in particular the $+1$ eigenspace of the stabilizers is also the 0 eigenspace of the modular variables,

$$\hat{T}_{j,0}|\psi\rangle = |\psi\rangle \Leftrightarrow \hat{x}_{j,[2\pi/l_j]}|\psi\rangle = 0. \tag{S20}$$

## III. FINITE-ENERGY GKP ENCODING

In the main Letter we have analyzed the case of a Gaussian envelope, $\hat{E}_\Delta = \exp\{-\Delta^2 \hat{n}\}$, and here we generalize this approach to trace-class functions of $\hat{n}$, $\hat{E}(\hat{n})$, which have an inverse $\hat{E}^{-1}$ such that $\hat{E}^{-1}\hat{E} = \mathbb{I}$. Since they are trace-class, envelope operators can also be written in the displacement operator basis,

$$\hat{E} = \int \frac{d^2\alpha}{\pi} \mathrm{Tr}[\hat{D}^\dagger(\alpha)\hat{E}]\hat{D}(\alpha). \tag{S21}$$

Using the relation $\langle n|\hat{D}(\alpha)|n\rangle = e^{-|\alpha|^2/2} L_n(|\alpha|^2)$, where $L_n(x)$ is the $n^{th}$ Laguerre polynomial, we get

$$\hat{E} = \sum_{n=0}^{\infty} E(n) \int \frac{d^2\alpha}{\pi} e^{\frac{-|\alpha|^2}{2}} L_n(|\alpha|^2) \hat{D}(\alpha). \tag{S22}$$

For the special case of the Gaussian envelope, the sum can be exactly evaluated

$$\hat{E}_\Delta = e^{-\Delta^2 \hat{n}},$$
$$= \frac{1}{\pi(1 - e^{-\Delta^2})} \int_{-\infty}^{\infty} d^2\alpha\, \hat{D}(\alpha) e^{-\frac{|\alpha|^2}{2\tanh(\Delta^2/2)}}. \tag{S23}$$

The first line of the above equation indicates that the effect of $\hat{E}_\Delta$ is to reduce the amplitude of the large Fock states, and the second line shows that this effect can also be understood as applying a coherent superposition of small displacements (when $\Delta \ll 1$). The destructive interference between the displaced GKP states effectively reduces the amplitude of large Fock states.

## A. States

We define generalized finite-energy states as

$$|\mu_E\rangle = \mathcal{N}_{E,\mu}\hat{E}(\hat{n})|\mu_0\rangle, \tag{S24}$$

where $\mathcal{N}_{E,\mu}$ is a normalization factor that depends on the exact form of $\hat{E}$. This factor can be found through

$$
\begin{aligned}
\langle \mu_E|\mu_E\rangle &= 1, \\
&= \mathcal{N}_{E,\mu}^2 \langle \mu_0|\hat{E}^\dagger \hat{E}|\mu_0\rangle, \\
&= \mathcal{N}_{E,\mu}^2 \sum_{n=0}^\infty \langle \mu_0|\hat{E}^\dagger|n\rangle\langle n|\hat{E}|\mu_0\rangle, \\
&= \mathcal{N}_{E,\mu}^2 \sum_{n=0}^\infty |E(n)|^2 |\langle n|\mu_0\rangle|^2,
\end{aligned}
\tag{S25}
$$

from which we obtain

$$\mathcal{N}_{E,\mu} = \left\{ \sum_{n=0}^\infty |E(n)|^2 |\langle n|\mu_0\rangle|^2 \right\}^{-1/2}. \tag{S26}$$

For a Gaussian envelope $\hat{E}_\Delta = \exp\{-\Delta^2 \hat{n}\}$ and a square lattice GKP, we get more specifically

$$
\begin{aligned}
\mathcal{N}_{\Delta,\mu} &= \left\{ \sum_{n=0}^\infty \sum_{j,j'\in\mathbb{Z}} e^{-2\Delta^2 n} \langle l_\square(j'+\mu/2)|n\rangle\langle n|l_\square(j+\mu/2)\rangle \right\}^{-1/2}, \\
&= \left\{ \frac{1}{\sqrt{\pi(1-e^{-2\Delta^2})}} \sum_{j,j'\in\mathbb{Z}} e^{\frac{-t_\square^2 t_\Delta}{4}(j+j'+\mu)^2} e^{\frac{-t_\square^2}{4t_\Delta}(j-j')^2} \right\}^{-1/2}, \\
&= \left\{ \frac{1}{\sqrt{\pi(1-e^{-2\Delta^2})}} \left[ \sum_{k_+,k_-\in\mathbb{Z}} e^{\frac{-t_\square^2 t_\Delta}{4}(2k_++\mu)^2} e^{\frac{-t_\square^2}{t_\Delta}k_-^2} + e^{\frac{-t_\square^2 t_\Delta}{4}(2k_++1+\mu)^2} e^{\frac{-t_\square^2}{4t_\Delta}(2k_--1)^2} \right] \right\}^{-1/2}, \\
&= \left\{ \frac{e^{-\pi t_\Delta \mu}}{\sqrt{\pi(1-e^{-2\Delta^2})}} \left[ \vartheta\left(i2t_\Delta\mu; i4t_\Delta\right)\vartheta\left(0; \frac{4i}{t_\Delta}\right) + e^{\frac{-\pi}{c_\Delta s_\Delta}}\vartheta\left(2it_\Delta(1-\mu); 4it_\Delta\right)\vartheta\left(\frac{2i}{t_\Delta}; \frac{4i}{t_\Delta}\right) \right] \right\}^{-1/2},
\end{aligned}
\tag{S27}
$$

where $\vartheta(z;\tau) = \sum_n \exp\{i\pi n^2\tau + i2\pi nz\}$ is Jacobi's theta function and we have defined $s_\Delta = \sinh\Delta^2$, $c_\Delta = \cosh\Delta^2$, $t_\Delta = \tanh\Delta^2$ for conciseness. Because of the envelope, the finite-energy codewords are not

orthogonal, and their overlap is given by

$$
\begin{aligned}
\langle 0_\Delta | 1_\Delta \rangle &= \mathcal{N}_{\Delta,0}\mathcal{N}_{\Delta,1} \sum_{n=0}^{\infty} \sum_{j,j'\in\mathbb{Z}} \mathrm{e}^{-2\Delta^2 n} \langle l_\square j' | n \rangle \langle n | l_\square (j+1/2) \rangle, \\
&= \frac{\mathcal{N}_{\Delta,0}\mathcal{N}_{\Delta,1}}{\sqrt{\pi(1-\mathrm{e}^{-2\Delta^2})}} \sum_{j,j'\in\mathbb{Z}} \mathrm{e}^{-\pi t_\Delta(j+j'+1/2)^2} \mathrm{e}^{\frac{-\pi}{t_\Delta}(j-j'-1/2)^2}, \\
&= \frac{\mathcal{N}_{\Delta,0}\mathcal{N}_{\Delta,1}}{\sqrt{\pi(1-\mathrm{e}^{-2\Delta^2})}} \left[ \sum_{k_+,k_-\in\mathbb{Z}} \mathrm{e}^{-\pi t_\Delta(2k_++1/2)^2} \mathrm{e}^{\frac{-\pi}{t_\Delta}(2k_--1/2)^2} + \mathrm{e}^{-\pi t_\Delta(2k_+-1/2)^2} \mathrm{e}^{\frac{-\pi}{t_\Delta}(2k_-+1/2)^2} \right], \\
&= \frac{\mathcal{N}_{\Delta,0}\mathcal{N}_{\Delta,1}\mathrm{e}^{-\frac{\pi}{4}\left(t_\Delta+\frac{1}{t_\Delta}\right)}}{\sqrt{\pi(1-\mathrm{e}^{-2\Delta^2})}} \left[ \vartheta\left(it_\Delta; i4t_\Delta\right)\vartheta\left(\frac{-i}{t_\Delta};\frac{4i}{t_\Delta}\right) + \vartheta\left(-it_\Delta; 4it_\Delta\right)\vartheta\left(\frac{i}{t_\Delta};\frac{4i}{t_\Delta}\right) \right].
\end{aligned}
\tag{S28}
$$

## B. Stabilizers

We define the finite-energy stabilizers through the similarity transformation induced by $\hat{E}$,

$$
\begin{aligned}
\hat{T}_{i,E} &= \hat{E}\hat{T}_{i,0}\hat{E}^{-1}, \\
&= \mathrm{e}^{i\hat{E}\hat{q}_j\hat{E}^{-1}}.
\end{aligned}
\tag{S29}
$$

Note that the above stabilizer can only be defined when $\hat{E}^{-1}$ exists, which is true only if $\hat{E}$ has support over all excitation numbers $n$. It is easy to check that the finite-energy stabilizers commute,

$$
\begin{aligned}
\left[\hat{T}_{1,E}, \hat{T}_{2,E}\right] &= \hat{T}_{1,E}\hat{T}_{2,E} - \hat{T}_{2,E}\hat{T}_{1,E}, \\
&= \hat{E}\hat{T}_{1,0}\hat{E}^{-1}\hat{E}\hat{T}_{2,0}\hat{E}^{-1} - \hat{E}\hat{T}_{2,0}\hat{E}^{-1}\hat{E}\hat{T}_{1,0}\hat{E}^{-1}, \\
&= \hat{E}\left(\hat{T}_{1,0}\hat{T}_{2,0} - \hat{T}_{2,0}\hat{T}_{1,0}\right)\hat{E}^{-1}, \\
&= 0.
\end{aligned}
\tag{S30}
$$

From the commutation relation of $\hat{a}$, we find $\hat{a}\hat{n} = (\hat{n}+1)\hat{a}$. By induction, we find $\hat{a}\hat{n}^p = (\hat{n}+1)^p\hat{a}$ for any $p \in \mathbb{N}$ and

$$
\hat{E}\hat{a}\hat{E}^{-1} = \hat{a}\hat{E}(\hat{n}-1)\hat{E}^{-1}(\hat{n}),
\tag{S31a}
$$

$$
\hat{E}\hat{a}^\dagger\hat{E}^{-1} = \hat{a}^\dagger\hat{E}(\hat{n}+1)\hat{E}^{-1}(\hat{n}),
\tag{S31b}
$$

from which we compute

$$
\begin{aligned}
\hat{E}\hat{q}_j\hat{E}^{-1} &= \alpha_j\hat{E}\hat{x}\hat{E}^{-1} + \beta_j\hat{E}\hat{p}\hat{E}^{-1}, \\
&= \alpha_j\hat{E}(\hat{n})\left(\frac{\hat{a}+\hat{a}^\dagger}{\sqrt{2}}\right)\hat{E}^{-1}(\hat{n}) - i\beta_j\hat{E}(\hat{n})\left(\frac{\hat{a}-\hat{a}^\dagger}{\sqrt{2}}\right)\hat{E}^{-1}(\hat{n}), \\
&= \hat{a}\frac{[\alpha_j-i\beta_j]\hat{E}(\hat{n}-1)\hat{E}^{-1}(\hat{n})}{\sqrt{2}} + \hat{a}^\dagger\frac{[\alpha_j+i\beta_j]\hat{E}(\hat{n}+1)\hat{E}^{-1}(\hat{n})}{\sqrt{2}}.
\end{aligned}
\tag{S32}
$$

Defining the functions

$$
\hat{c}(\hat{n}) = \frac{\hat{E}(\hat{n}-1)\hat{E}^{-1}(\hat{n}) + \hat{E}(\hat{n}+1)\hat{E}^{-1}(\hat{n})}{2},
\tag{S33a}
$$

$$
\hat{s}(\hat{n}) = \frac{\hat{E}(\hat{n}-1)\hat{E}^{-1}(\hat{n}) - \hat{E}(\hat{n}+1)\hat{E}^{-1}(\hat{n})}{2},
\tag{S33b}
$$

we can conveniently write Eq. (S32) as the sum of a Hermitian and an anti-Hermitian term,

$$
\begin{aligned}
\hat{E}\hat{q}_j\hat{E}^{-1} &= [\alpha_j\hat{x} + \beta_j\hat{p}]\,\hat{c}(\hat{n}) + i[\alpha_j\hat{p} - \beta_j\hat{x}]\,\hat{s}(\hat{n}), \\
&= \hat{q}_j\hat{c}(\hat{n}) + i\hat{q}_{j,\perp}\hat{s}(\hat{n}).
\end{aligned}
\tag{S34}
$$

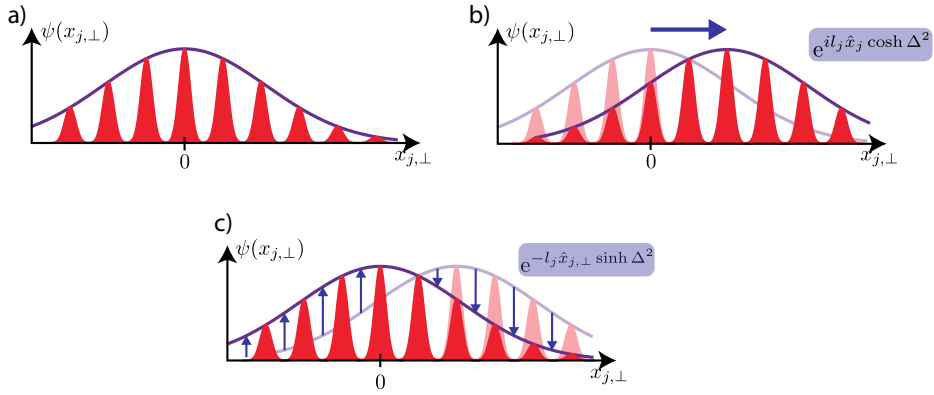

FIG. S2. Effect of the finite-energy stabilizer $\hat{T}_{j,\Delta} = \mathrm{e}^{r_j}\mathrm{e}^{-\hat{q}_{j,\perp}\sinh\Delta^2}\mathrm{e}^{i\hat{q}_j\cosh\Delta^2}$ (blue) on a GKP wavefunction (red) with a Gaussian envelope (purple). The initial wavefunction (a) is translated by $l_j\cosh\Delta^2$ (b) and then rescaled by an exponential (c).

The general finite-energy stabilizers are given by

$$\hat{T}_{j,E} = \mathrm{e}^{i\hat{q}_j\hat{c}(\hat{n}) - \hat{q}_{j,\perp}\hat{s}(\hat{n})}.\tag{S35}$$

This relation shows that it is possible to define exact finite-energy stabilizers for a broad class of envelopes. However, for general functions of $\hat{n}$, the operators $\hat{c}$ and $\hat{s}$ are likely to be complicated, making the stabilization of general GKP states challenging.

In the particular case of the Gaussian envelope, $\hat{E}(\hat{n}) = \hat{E}_\Delta = \exp\{-\Delta^2\hat{n}\}$, the dependence on $\hat{n}$ disappears, and we find the simpler expression

$$\hat{E}_\Delta \hat{q}_j \hat{E}_\Delta^{-1} = \hat{q}_j\cosh\Delta^2 + i\hat{q}_{j,\perp}\sinh\Delta^2,\tag{S36}$$

from which equations 2a) and 2b) from the main Letter follow. To gain further understanding in the finite-energy stabilizers for the Gaussian envelope, we rewrite them as a combination of a unitary translation operator and a rescaling operator

$$\begin{aligned}\hat{T}_{j,\Delta} &= \mathrm{e}^{i[\hat{q}_j\cosh\Delta^2 + i\hat{q}_{j,\perp}\sinh\Delta^2]},\\&= \mathrm{e}^{r_j}\times\mathrm{e}^{-\hat{q}_{j,\perp}\sinh\Delta^2}\times\mathrm{e}^{i\hat{q}_j\cosh\Delta^2},\end{aligned}\tag{S37}$$

with $r_j = \cosh\Delta^2\sinh\Delta^2 l_j^2/2 \in \mathbb{R}$. To derive the above equation, we used the BCH formula $\exp\{\hat{A}+\hat{B}\} = \exp\{\hat{A}\}\exp\{\hat{B}\}\exp\{-[\hat{A},\hat{B}]/2\}$, which is exact when $[\hat{A},\hat{B}]$ commute with both $\hat{A}$ and $\hat{B}$. In the limit $\Delta \to 0$ the ideal stabilizers are recovered, justifying the choice of subscript for the ideal code. The effect of $\hat{T}_{j,\Delta}$ in the $\hat{q}_{j,\perp}$ eigenbasis, as schematically illustrated in Fig. S2, is to first displace the wavefunction and then to rescale the envelope such that the initial GKP wavefunction is recovered. Alternatively, we can write the finite-energy stabilizers using the squeezing operator,

$$\hat{T}_{j,\Delta} = \hat{S}^\dagger(\xi)\mathrm{e}^{il_j\mathrm{e}^{-i\theta}\sqrt{2\cosh\Delta^2\sinh\Delta^2}\hat{a}}\hat{S}(\xi),\tag{S38}$$

with $\xi = \mathrm{e}^{2i\theta}\ln\sqrt{\tanh\Delta^2}$, $\alpha_j + i\beta_j = l_j\mathrm{e}^{i\theta}$. While finite-energy GKP states are traditionally understood by starting from superposition of infinitely squeezed states (e.g. Eq. (S24)), the equation above shows that one can also start from the reverse limit, taking superposition of coherent states and gradually increasing the squeezing. In the main Letter, we studied the case $\hat{T}_{x,\Delta} = \exp\{il_\square\hat{x}\}$ for which $\theta = 0$. However, Eq. (S38) shows that for any lattice, the eigenstates of the (Gaussian) finite-energy stabilizers are squeezed coherent states. Consequently, the finite-energy states defined in Eq. (S24) are equivalent to a *superposition* of finitely squeezed states when $\hat{E}(\hat{n}) = \exp\{-\Delta^2\hat{n}\}$.

We can understand the finite-energy stabilizers in a third way from the logarithm of the stabilizer,

$$\ln\hat{T}_{j,\Delta} = il_j\sqrt{2\cosh\Delta^2\sinh\Delta^2}\left(\frac{\frac{\hat{x}_{j,[2\pi/(l_j\cosh\Delta^2)]}}{\sqrt{\tanh\Delta^2}} + i\hat{x}_{j,\perp}\sqrt{\tanh\Delta^2}}{\sqrt{2}}\right).\tag{S39}$$

We define the last part as an operator

$$\hat{d}_{j,\Delta} = \frac{\frac{\hat{x}_{j,[m_j]}}{\sqrt{\tanh \Delta^2}} + i\hat{x}_{j,\perp}\sqrt{\tanh \Delta^2}}{\sqrt{2}}, \tag{S40}$$

where we have defined $m_j = 2\pi/(l_j \cosh \Delta^2)$ for conciseness. This operator is similar to a squeezed annihilation operator $\hat{S}^\dagger(\xi)\hat{a}\hat{S}(\xi)$, except that the $\hat{x}_j$ variable is taken as a modular variable due to the multivalued nature of the complex logarithm. From this squeezed annihilation-like operator, the stabilizer condition is expressed as

$$\hat{T}_{j,\Delta}|\psi\rangle = |\psi\rangle \Leftrightarrow \hat{d}_{j,\Delta}|\psi\rangle = 0. \tag{S41}$$

### C. Logical Pauli Operators

We define the logical Pauli operators in an analogous way to the generators of the stabilizer group,

$$\hat{X}_E = \hat{E}\hat{X}_0\hat{E}^{-1}, \\ \hat{Z}_E = \hat{E}\hat{Z}_0\hat{E}^{-1}, \tag{S42}$$

and using Eq. (S34) we get

$$\hat{X}_E = e^{\frac{i\hat{q}_1\hat{c}(\hat{n})-\hat{q}_{1,\perp}\hat{s}(\hat{n})}{2}}, \\ \hat{Z}_E = e^{\frac{i\hat{q}_2\hat{c}(\hat{n})-\hat{q}_{2,\perp}\hat{s}(\hat{n})}{2}}, \tag{S43}$$

with $\hat{Y}_E = -i\hat{Z}_E\hat{X}_E$. Since the finite-energy logical operators are defined from a similarity transformation, we have that for any $\hat{A}_E, \hat{B}_E \in \{\hat{X}_E, \hat{Y}_E, \hat{Z}_E, \hat{T}_{1,E}, \hat{T}_{2,E}\}$,

$$\begin{aligned} \left[\hat{A}_E, \hat{B}_E\right] &= \hat{A}_E\hat{B}_E - \hat{B}_E\hat{A}_E, \\ &= \hat{E}\hat{A}_0\hat{E}^{-1}\hat{E}\hat{B}_0\hat{E}^{-1} - \hat{E}\hat{B}_0\hat{E}^{-1}\hat{E}\hat{A}_0\hat{E}^{-1}, \\ &= \hat{E}\left(\hat{A}_0\hat{B}_0 - \hat{B}_0\hat{A}_0\right)\hat{E}^{-1}, \\ &= \hat{E}\left[\hat{A}_0, \hat{B}_0\right]\hat{E}^{-1}. \end{aligned} \tag{S44}$$

Consequently, the finite-energy operators obey the same commutation relation as their ideal counterparts.

In the case of the Gaussian envelope, we have the simpler expressions for the logical Pauli operators

$$\hat{X}_\Delta = e^{\frac{i\hat{q}_1 \cosh \Delta^2 - \hat{q}_{1,\perp} \sinh \Delta^2}{2}}, \\ \hat{Z}_\Delta = e^{\frac{i\hat{q}_2 \cosh \Delta^2 - \hat{q}_{2,\perp} \sinh \Delta^2}{2}}. \tag{S45}$$

## IV. RESERVOIR ENGINEERING

In this section, we provide a more detailed derivation of the circuits in Fig. 2 of the main Letter. Concisely, we adapt the discretization procedure of collisional models of dissipation to the finite-energy GKP case, and we refer the reader to Refs. [3–6] and references therein for more details. We focus on the case of a Gaussian envelope, $\hat{E}_\Delta$, and leave the stabilization of more general envelopes for future work.

### A. Continuous model

We first discuss a continuous, Markovian model of dissipation, assuming that the oscillator-bath interaction is described by the Hamiltonian

$$\hat{H}_{j,\Delta}(t) = \sqrt{\Gamma_j}[\hat{d}_{j,\Delta}\hat{b}_t^\dagger + \hat{d}_{j,\Delta}^\dagger\hat{b}_t], \tag{S46}$$

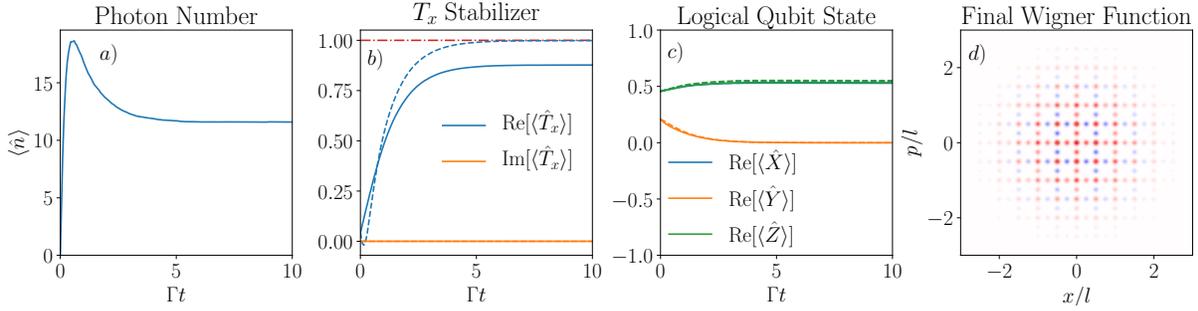

FIG. S3. Continuous dynamics generated by the master equation Eq. (S50) for a square lattice GKP, starting from vacuum with $\Gamma_1 = \Gamma_2 = \Gamma$ and $\Delta = 0.2$. a) Excitation number as a function of time. b) Real and imaginary parts of the stabilizer $\hat{T}_x$ as a function of time. c) Real part of the logical Pauli operators. In panels b) and c), full lines correspond to the ideal operators (index 0), while the dashed lines refer to the finite-energy operators (index $\Delta$). d) Wigner function of the final state at $\Gamma t = 10$.

where the bath operators obey $[\hat{b}_t, \hat{b}_{t'}^\dagger] = \delta(t-t')$, with $\delta(x)$ the Dirac delta function. The evolution generated by that Hamiltonian is then given by

$$\hat{U}(t, t_0) = \mathcal{T} e^{-i \int_{t_0}^t d\tau \hat{H}_{j,\Delta}(\tau)},$$ (S47)

with $\mathcal{T}$ the time-ordering operator. Taking a zero temperature bath with average values

$$\begin{aligned} \langle \hat{b}_t^\dagger \hat{b}_{t'} \rangle &= 0, \\ \langle \hat{b}_t \hat{b}_{t'} \rangle &= 0, \\ \langle \hat{b}_t^\dagger \hat{b}_{t'}^\dagger \rangle &= 0, \end{aligned}$$ (S48)

and following standard methods [1], the bath can be traced out which yields a master equation

$$\dot{\rho} = \Gamma_j \mathcal{D}\left[\hat{d}_{j,\Delta}\right] \rho,$$ (S49)

where $\mathcal{D}[\hat{O}]\bullet = \hat{O} \bullet \hat{O}^\dagger - \{\hat{O}^\dagger \hat{O}, \bullet\}/2$ is the standard dissipation superoperator. With a second, independent bath for the other stabilizer, the full finite-energy GKP manifold is stabilized. The equivalent master equation is given by

$$\dot{\rho} = \Gamma_1 \mathcal{D}\left[\hat{d}_{1,\Delta}\right] \rho + \Gamma_2 \mathcal{D}\left[\hat{d}_{2,\Delta}\right] \rho.$$ (S50)

The time evolution of an oscillator initialized in vacuum and stabilized by Eq. (S50) is shown in Fig. S3 for a square lattice. The modular variable $\hat{x}_{[l/2\cosh\Delta^2]}$ is written using its Fourier series Eq. (S18) which we truncate at $j = 30$. In panel a), the excitation number as a function of time is shown. There is no reason why the excitation number should evolve monotically under Eq. (S50) and the excitation number makes an initial overshoot before stabilizing to its long time steady state value. Panel b) shows the time-evolution of the real (blue) and imaginary (orange) parts of the average value for the ideal $\hat{T}_{x,0}$ stabilizers (full lines) and finite-energy $\hat{T}_{x,\Delta}$ stabilizers (dashed lines). The red dashed-dotted line at a fixed value of 1 is a guide for the eye. While the average of the ideal stabilizer saturates at $\approx 0.88 < 1$, the finite-energy stabilizers stabilizes to the desired $+1$ value. The time-evolution for the $\hat{T}_p$ stabilizer is similar to Fig. S3b). In panel c), the logical Pauli operator average values are shown. As in panel b), full lines refer to the ideal operators, while the dashed line refer to the finite-energy operators. The fact that these average values do not change in the long time limit indicates that the bath does not induce decoherence in the logical manifold. Since the projection of the vacuum state onto the logical manifold does not yield a pure qubit state, the stabilized logical GKP qubit is in a mixed state. The Wigner function for the final state at $\Gamma t = 10$ is shown in panel d).

While the approach described above leads to the autonomous stabilization of finite-energy GKP states, interactions of the form described by Eq. (S46) seem challenging to realize in practice. We therefore look for simplifications that make this approach more realistic.

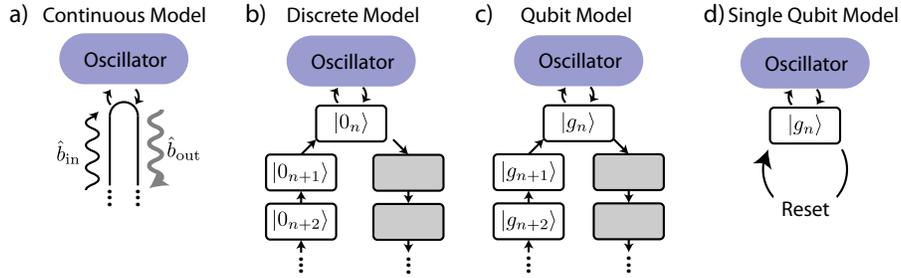

FIG. S4. Schematic representation of the discretization procedure. a) Continuous model with a vacuum input field and an output field "carrying away" excitations of the GKP state. b) Discretization of the bath modes over slices of time $\delta t$. c) In the limit that the number of excitations in the $|\phi_n\rangle$ states is smaller than 1, the bath modes can be replaced by qubits. d) Since the qubit states are not reused after the interaction, the ensemble of qubits can be replaced by a single one that is reset between each step.

## B. Discretized model

One interpretation of the Hamiltonian Eq. (S46) is that the oscillator interacts with a different bath mode at each time $t$. A good approximation is then to consider that each bath mode extends over finite slices of time $\delta t$. This can be done by replacing $\hat{b}_t \to \hat{b}_n/\sqrt{\delta t}$ with $[\hat{b}_n, \hat{b}_{n'}^\dagger] = \delta_{n,n'}$, with $\delta_{x,y}$ denoting the Kronecker delta, $n \in \mathbb{Z}$ and $t = n\delta t$. In order to match the average values Eq. (S48), all bath modes are initialized in vacuum, which we denote $|0_n\rangle$. In this discretized picture, the time evolution operator Eq. (S47) is replaced by the "conveyor belt" model illustrated in Fig. S4b, which translates into equation form to

$$\hat{U}(t, t_0) = \Pi_{n=0}^T \mathrm{e}^{-i\sqrt{\Gamma_j \delta t}(\hat{d}_{j,\Delta}\hat{b}_n^\dagger + \hat{d}_{j,\Delta}^\dagger \hat{b}_n)},$$
$$= \Pi_{n=0}^T \hat{U}_n,$$

(S51)

where $t - t_0 = T\delta t$, $T \in \mathbb{Z}$, and the multiplication is time-ordered. In the limit $\delta t \to 0$, the continuous model is recovered. For small slices of time and an initial oscillator-bath state $|\psi\rangle \otimes |0_n\rangle = |\psi, 0_n\rangle$, we have

$$\langle\psi, 0_n|\hat{U}_n^\dagger \hat{b}_n^\dagger \hat{b}_n \hat{U}_n|\psi, 0_n\rangle \approx \Gamma_j \delta t \, \langle\psi|\hat{d}_{j,\Delta}^\dagger \hat{d}_{j,\Delta}|\psi\rangle + \mathcal{O}(\Gamma_j^2 \delta t^2),$$

(S52)

which means that the number of excitations in the bath is proportional to $\Gamma_j \delta t$.

## C. Qubit model

If $\Gamma_j \delta t$ is small enough, then after interacting with the oscillator, the $n^{th}$ bath mode contains on average less than one excitation. In that regime, the bath modes can be equivalently replaced by qubits, $\hat{b}_n \to (\hat{\sigma}_{x,n} + i\hat{\sigma}_{y,n})/\sqrt{2}$, with $\hat{\sigma}_{\alpha,n}$ denoting the Pauli matrices of the $n^{th}$ qubit mode, $\alpha \in \{x, y, z\}$. We denote the ground and excited state of the ancilla qubit by $|g\rangle$ and $|e\rangle$, respectively, and the vacuum bath modes are mapped to $|0_n\rangle \to |g_n\rangle$ as illustrated in Fig. S4c. The commutation relation of the bath operators are correspondingly mapped to $[\hat{b}_n, \hat{b}_n^\dagger] = 1 \to [(\hat{\sigma}_{x,n} + i\hat{\sigma}_{y,n})/\sqrt{2}, (\hat{\sigma}_{x,n} - i\hat{\sigma}_{y,n})/\sqrt{2}] = \hat{\sigma}_{z,n}$. In the limit that the qubit is very weakly populated, then $\langle\hat{\sigma}_{z,n}\rangle \approx 1$ and the commutation relations for the bath are almost unchanged. In looser terms, the neighborhood of any point on a Bloch sphere looks like the flat phase space of an oscillator for small enough displacements. In the qubit model, the time-evolution is replaced by

$$\hat{U}(t, t_0) = \Pi_n \mathrm{e}^{-i\sqrt{\frac{\Gamma_j \delta t}{2}}[\hat{d}_{j,\Delta}(\hat{\sigma}_{x,n} - i\hat{\sigma}_{y,n}) + \hat{d}_{j,\Delta}^\dagger(\hat{\sigma}_{x,n} + i\hat{\sigma}_{y,n})]},$$
$$= \Pi_n \mathrm{e}^{-i\sqrt{\frac{\Gamma_j \delta t}{\tanh \Delta^2}}[\hat{x}_{j,[m_j]}\hat{\sigma}_{x,n} + \hat{x}_{j,\perp}\hat{\sigma}_{y,n}\tanh \Delta^2]}.$$

(S53)

In the limit $\delta t \to 0$, the unitary evolution Eq. (S53) is equivalent to the continuous evolution Eq. (S47). Moreover, in the input-output model considered above, the qubits carry away the entropy from the oscillator and are not used again as schematically illustrated in Fig. S4c. We can therefore replace the qubit ensemble

by a single ancilla that is reset between each step, Fig. S4d. Removing the $n$ index from Eq. (S53), the $+1$ eigenstate of the $\hat{T}_{j,\Delta}$ operator is stabilized by repeated application of

$$\hat{U}_{\text{target}} = e^{-i\sqrt{\frac{\Gamma_j \delta t}{\tanh \Delta^2}}\left[\hat{x}_{j,[m_j]}\hat{\sigma}_x + \hat{x}_{j,\perp}\hat{\sigma}_y \tanh \Delta^2\right]},$$

$$\hat{U}(t,t_0) = \Pi_n \hat{U}_{\text{target}}, \tag{S54}$$

where the qubit is initialized in $|g\rangle$ and reset at the end of each step. While it is conceivable that the above unitary could be implemented directly, coupling to a modular variable is challenging and we look for further simplifications of Eq. (S54). We find three such simplifications, which give rise to three protocols stabilizing the desired manifold.

### 1. Sharpen-Trim Protocol

A first-order Trotter decompositon of the target unitary,

$$e^{\hat{A}+\hat{B}} \approx e^{\hat{A}}e^{\hat{B}} - \frac{1}{2}[\hat{A},\hat{B}] + ...,$$

$$e^{\hat{A}+\hat{B}} \approx e^{\hat{B}}e^{\hat{A}} + \frac{1}{2}[\hat{A},\hat{B}] + ..., \tag{S55}$$

yields the two approximations

$$\hat{U}_{\text{target}} \approx e^{-i\sqrt{\Gamma_j \delta t \tanh \Delta^2}\hat{x}_{j,\perp}\hat{\sigma}_y}e^{-i\sqrt{\frac{\Gamma_j \delta t}{\tanh \Delta^2}}\hat{x}_{j,[m_j]}\hat{\sigma}_x} + \mathcal{O}(\Gamma_j \delta t),$$

$$\approx e^{-i\sqrt{\frac{\Gamma_j \delta t}{\tanh \Delta^2}}\hat{x}_{j,[m_j]}\hat{\sigma}_x}e^{-i\sqrt{\Gamma_j \delta t \tanh \Delta^2}\hat{x}_{j,\perp}\hat{\sigma}_y} + \mathcal{O}(\Gamma_j \delta t). \tag{S56}$$

One of the defining features of the modular quadrature $\hat{x}_{j,[m]}$ is that it is invariant under translations by $m$, in other words $e^{im\hat{x}_{j,\perp}}\hat{x}_{j,[m]}e^{-im\hat{x}_{j,\perp}} = \hat{x}_{j,[m]}$. Instead of imposing this constraint on the qubit-oscillator coupling, what we suggest is to impose a similar constraint on the *whole* unitary. This permits the replacement $\hat{x}_{j,[m_j]} \rightarrow \hat{x}_j$ and, with the choice

$$\Gamma_j \delta t = \tanh \Delta^2 \left(\frac{l_j \cosh \Delta^2}{2}\right)^2, \tag{S57}$$

we obtain

$$\hat{U}_{(i)}^{(ST)} = e^{-i\frac{l_j \sinh \Delta^2}{2}\hat{x}_{j,\perp}\hat{\sigma}_y}e^{-i\frac{l_j \cosh \Delta^2}{2}\hat{x}_j\hat{\sigma}_x},$$

$$\hat{U}_{(ii)}^{(ST)} = e^{-i\frac{l_j \cosh \Delta^2}{2}\hat{x}_j\hat{\sigma}_x}e^{-i\frac{l_j \sinh \Delta^2}{2}\hat{x}_{j,\perp}\hat{\sigma}_y}, \tag{S58}$$

which both obey

$$e^{im_j\hat{x}_{j,\perp}}\hat{U}_{(\alpha)}^{(ST)}e^{-im_j\hat{x}_{j,\perp}} = -\hat{U}_{(\alpha)}^{(ST)}, \tag{S59}$$

with $\alpha \in \{i,ii\}$. Since $\hat{U}_{(i)}^{(ST)}$ and $\hat{U}_{(ii)}^{(ST)}$ anti-commute with the translation operator, they apply a -1 sign to every other peak of the GKP state in the $\hat{x}_{j,\perp}$ direction. In other words, $\hat{U}_{(i)}^{(ST)}$ and $\hat{U}_{(ii)}^{(ST)}$ apply a logical Pauli operation on top of stabilizing the logical manifold. This deterministic operation can be taken care of through Pauli frame updating. Defining

$$\epsilon_j = \sinh \Delta^2 l_j, \tag{S60}$$

we obtain the simpler unitaries

$$\hat{U}_{(i)}^{(ST)} = e^{-i\frac{\epsilon_j}{2}\hat{x}_{j,\perp}\hat{\sigma}_y}e^{-i\frac{l_j \cosh \Delta^2}{2}\hat{x}_j\hat{\sigma}_x},$$

$$\hat{U}_{(ii)}^{(ST)} = e^{-i\frac{l_j \cosh \Delta^2}{2}\hat{x}_j\hat{\sigma}_x}e^{-i\frac{\epsilon_j}{2}\hat{x}_{j,\perp}\hat{\sigma}_y}. \tag{S61}$$

While both unitaries approximate Eq. (S54), they lead to non-negligible errors if taken alone. For a better approximation, we alternate between the unitaries $(i)$ and $(ii)$, which cancels the leading error term $[\hat{A}, \hat{B}]$ in Eq. (S55). In practice, only applying the "sharpening" unitary $(i)$ leads to a steady increase of the excitation number and the GKP envelope. In contrast, only applying the "trimming" unitary $(ii)$ leads to a decrease in excitation number. In short, $\hat{U}_{(i)}^{(ST)}$ stabilizes the GKP grid and $\hat{U}_{(ii)}^{(ST)}$ stabilizes the envelope. Note that $\delta t$ is usually set by experimental parameters, and the choice Eq. (S57) fixes $\Gamma_j \propto \Delta^2$. Since $l_j$ and $\hat{x}_j$ are fixed by the choice of GKP lattice, the only free parameter above is $\epsilon_j$, which directly sets the GKP size $\Delta$. Moreover, assuming that the individual parts of the unitary are generated by some fixed Hamiltonian of the form $\hat{H} \propto \hat{x}_j \hat{\sigma}_x$, we remark that only the completed interaction obeys the condition Eq. (S59). This means that qubit errors during the unitary can lead to the environment learning some information about the non-modularity of $\hat{x}_j$ and to logical errors in the GKP manifold.

As mentioned in the main Letter, we choose to express the target unitaries in terms of a controlled displacement operation $C\hat{D}(\beta) = \exp\{(\beta \hat{a}^\dagger - \beta^* \hat{a})\hat{\sigma}_z/2\sqrt{2}\}$. Using the relations $\hat{\sigma}_x = \hat{H}\hat{\sigma}_z\hat{H}$ and $\hat{\sigma}_y = \hat{R}_z(\pi/2)\hat{H}\hat{\sigma}_z\hat{H}\hat{R}_z^\dagger(\pi/2)$, where $\hat{H}$ is the Hadamard gate and $\hat{R}_\alpha(\theta) = \exp\{-i\theta\hat{\sigma}_\alpha/2\}$, we obtain the unitaries

$$\hat{U}_{(i)}^{(ST)} = \hat{R}_z(\pi/2)\hat{H}e^{-i\frac{\epsilon_j}{2}\hat{x}_{j,\perp}\hat{\sigma}_z}\hat{H}\hat{R}_z^\dagger(\pi/2)\hat{H}e^{-i\frac{l_j \cosh\Delta^2}{2}\hat{x}_j\hat{\sigma}_z}\hat{H},$$

$$\hat{U}_{(ii)}^{(ST)} = \hat{H}e^{-i\frac{l_j \cosh\Delta^2}{2}\hat{x}_j\hat{\sigma}_z}\hat{H}\hat{R}_z(\pi/2)\hat{H}e^{-i\frac{\epsilon_j}{2}\hat{x}_{j,\perp}\hat{\sigma}_z}\hat{H}\hat{R}_z^\dagger(\pi/2).$$

(S62)

Since the qubit is reset after the unitary, we can ignore the last qubit operations. Moreover, the first qubit operations can be included in the ancilla state preparation, now of the state $|+\rangle = (|g\rangle + |e\rangle)/\sqrt{2}$. Finally, including the middle Hadamards into the qubit rotation and recalling the definitions $\hat{x}_j = (\alpha_j \hat{x} + \beta_j \hat{p})/l_j$ and $\alpha_j + i\beta_j = l_j e^{i\theta_j}$, we get

$$\hat{U}_{(i)}^{(ST)} = C\hat{D}\left(\epsilon_j e^{i\theta_j}\right) \times \hat{R}_x^\dagger(\pi/2) \times C\hat{D}\left(-il_j e^{i\theta_j} \cosh\Delta^2\right),$$

$$\hat{U}_{(ii)}^{(ST)} = C\hat{D}\left(-il_j e^{i\theta_j} \cosh\Delta^2\right) \times \hat{R}_x(\pi/2) \times C\hat{D}\left(\epsilon_j e^{i\theta_j}\right),$$

(S63)

which are illustrated in Fig. S6a. For the $\hat{T}_{x,\Delta}$ stabilizer of the square lattice, we obtain the circuits in Fig. 2a of the main Letter.

In order for the qubit model to correctly approximate the discretized model of dissipation, the number of excitations transfered to the bath modes in one time step should be much smaller than 1. This imposes not only the condition of small time steps $\Gamma_j \delta \ll 1$, but also $\langle\psi|\hat{d}_{j,\Delta}^\dagger \hat{d}_{j,\Delta}|\psi\rangle \ll 1$, see Eq. (S52). This condition can fail if the state $|\psi\rangle$ is displaced by a large amount in the $\hat{x}_{j,\perp}$ direction. In practice, this means that for the ST protocol

$$e^{i\frac{2\pi}{\epsilon_j}\hat{x}_j}\hat{U}_{(\alpha)}^{(ST)}e^{-i\frac{2\pi}{\epsilon_j}\hat{x}_j} = -\hat{U}_{(\alpha)}^{(ST)},$$

(S64)

and so the ST protocols stabilizes a superlattice of lattice constant $2\pi/\epsilon_j$. Since the protocol is already designed to protect against small displacement errors, it also protects against displacement of order $\sim 1/\epsilon_j$, and the state stays confined within the central unit cell of the superlattice. Although in principle other sites of that superlattice could become populated, we did not observe any numerical evidence that oscillator errors could cause this type of tunneling for the ST protocol.

As shown in Fig. S5 and in the absence of errors, the last controlled-displacement of a protocol can be equivalently replaced by a qubit measurement followed by classical feedback. In order to highlight this relation, we chose a symmetric definition of the classical feedback, applying a displacement $\hat{D}(\pm\beta/2)$ if the qubit is measured in $|g\rangle$ or $|e\rangle$. Doing the replacement shown in Fig. S5 for the two circuits of the Sharpen-Trim protocol yields the same circuits that were introduced in Ref. [7]. There, the feedback value $\epsilon_{(i)}$ (which they call $\delta$) was optimized experimentally as well as the trimming value $\epsilon_{(ii)}$ (which they call $\epsilon$). Two natural questions that arise from the experiment of Ref. [7] are (expressed in the language of the present Letter):

- What is the optimal value of $\epsilon_{(i)}$ given a fixed $\epsilon_{(ii)}$ ?

- Are all values of $\epsilon_{(ii)}$ allowed ?

The results presented in this section provide an answer to both, with $\epsilon_{(i)} = \epsilon_{(ii)}$ being optimal and all values of $\epsilon_{(ii)}$ that are small enough being allowed. However, as discussed in the main Letter and below in Sect. VII A 3, there is a GKP size (set by $\epsilon$) that optimizes the lifetime of the information.

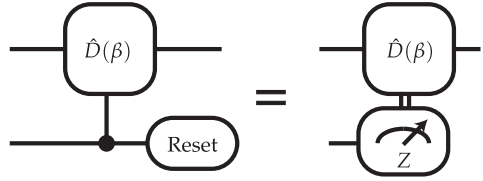

FIG. S5. The last controlled displacement of an autonomous protocol (left) can be modified into a qubit measurement followed by classical feedback (right). In the absence of errors, the two circuits are equivalent.

### 2. small-Big-small Protocol

We obtained the Sharpen-Trim protocol from a first order Trotter decomposition of the target unitary Eq. (S54). A better approximation of the target unitary is obtained from a second order Trotter decomposition

$$e^{\hat{A}+\hat{B}} \approx e^{\frac{\hat{A}}{2}} e^{\hat{B}} e^{\frac{\hat{A}}{2}}, \tag{S65}$$

and the two different choices for $\hat{A}$ and $\hat{B}$ result in the small-Big-small (sBs) or Big-small-Big (BsB) protocols. First, we consider the choice of $\hat{A} \propto \hat{x}_{j,\perp}$,

$$\hat{U}_{\text{target}} \approx e^{-i\sqrt{\Gamma_j \delta t \tanh \Delta^2} \frac{\hat{x}_{j,\perp} \hat{\sigma}_y}{2}} e^{-i\sqrt{\frac{\Gamma_j \delta t}{\tanh \Delta^2}} \hat{x}_{j,[m_j]} \hat{\sigma}_x} e^{-i\sqrt{\Gamma_j \delta t \tanh \Delta^2} \frac{\hat{x}_{j,\perp} \hat{\sigma}_y}{2}} + \mathcal{O}\left[(\Gamma_j \delta t)^{3/2}\right]. \tag{S66}$$

Replacing $\hat{x}_{j,[m_j]} \to \hat{x}_j$ and imposing the same anti-commutation condition as in Eq. (S59) leads to the same choice of $\Gamma_j$ as for the ST protocol, Eq. (S57). With the same definition for $\epsilon_j$ as in Eq. (S60), we get

$$\hat{U}^{(sBs)} = e^{-i\frac{\epsilon_j}{4} \hat{x}_{j,\perp} \hat{\sigma}_y} e^{-i\frac{l_j \cosh \Delta^2}{2} \hat{x}_j \hat{\sigma}_x} e^{-i\frac{\epsilon_j}{4} \hat{x}_{j,\perp} \hat{\sigma}_y}. \tag{S67}$$

Expressing the above unitary in terms of controlled displacement, we get the circuits in Fig. S6c.

Because of the approximation of replacing the bath with qubits, this protocol too stabilizes a large superlattice on top of the desired GKP lattice. However, because of the one-half factor in $\hat{A}/2$, individual parts of the unitary anti-commute with displacements of $4\pi/\epsilon_j$ instead of $2\pi/\epsilon_j$. Consequently, the sBs protocol stabilizes a superlattice which is twice as large as for the ST protocol. In our numerical simulations with oscillator errors, we observed no evidence of tunneling between sites of this larger superlattice.

### 3. Big-small-Big Protocol

From the decomposition Eq. (S65), we can also choose $\hat{A} \propto \hat{x}_{j,[m_j]}$, which leads to

$$\hat{U}_{\text{target}} \approx e^{-i\sqrt{\frac{\Gamma_j \delta t}{4 \tanh \Delta^2}} \hat{x}_{j,[m_j]} \hat{\sigma}_x} e^{-i\sqrt{\Gamma_j \delta t \tanh \Delta^2} \hat{x}_{j,\perp} \hat{\sigma}_y} e^{-i\sqrt{\frac{\Gamma_j \delta t}{4 \tanh \Delta^2}} \hat{x}_{j,[m_j]} \hat{\sigma}_x} + \mathcal{O}\left[(\Gamma_j \delta t)^{3/2}\right]. \tag{S68}$$

In contrast to the ST and sBs protocols, the unitary above contains two interactions with the modular variable $\hat{x}_{j,[m_j]}$. Replacing $\hat{x}_{j,[m_j]} \to \hat{x}_j$ and imposing the condition that each of these interactions anti-commute with the translation operator $e^{-im_j \hat{x}_{j,\perp}}$, we get

$$\hat{U}^{(BsB)} = e^{-i\frac{l_j \cosh \Delta^2}{2} \hat{x}_{j,[m_j]} \hat{\sigma}_x} e^{-i\epsilon_j \hat{x}_{j,\perp} \hat{\sigma}_y} e^{-i\frac{l_j \cosh \Delta^2}{2} \hat{x}_{j,[m_j]} \hat{\sigma}_x}, \tag{S69}$$

where we used the same definition for $\epsilon_j$, Eq. (S60), as in the two other protocols. Note that the completed BsB unitary *commutes* with the translation operator,

$$e^{im_j \hat{x}_{j,\perp}} \hat{U}^{(BsB)} e^{-im_j \hat{x}_{j,\perp}} = \hat{U}^{(BsB)}. \tag{S70}$$

which means that the BsB protocol does not apply a logical operation in the GKP manifold contrary to the ST and sBs protocols.

Similarly to the previous two protocols, the BsB protocol stabilizes a superlattice on top of the GKP lattice. However, the choice of $\Gamma_j$ here leads to a superlattice twice smaller, $\pi/\epsilon_j$, than the superlattice for the ST protocol. The effects of this superlattice are discussed more in detail below.

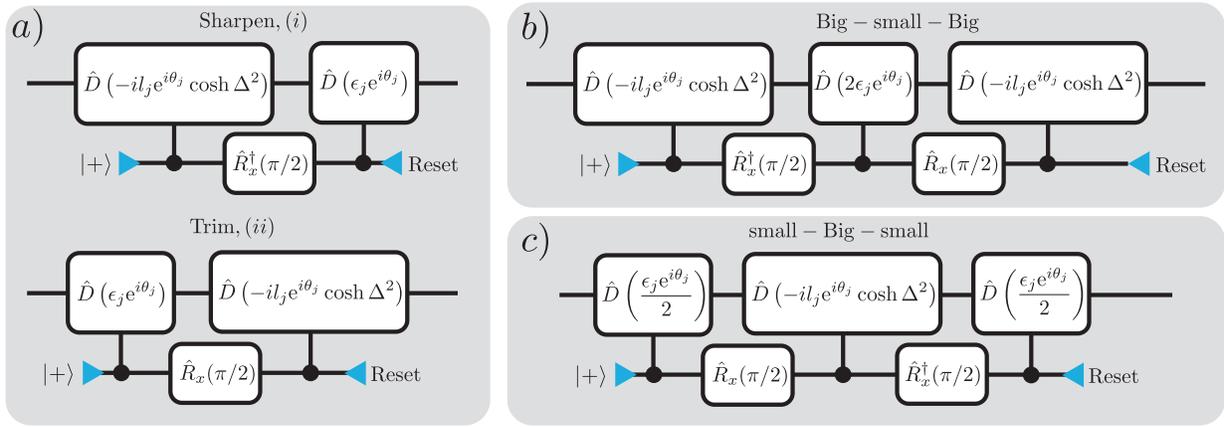

FIG. S6. Generalized version of the circuits presented in the main Letter.

| Protocols | Short-hand | Cooling rate, $\Gamma_j \delta t$ | Superlattice constant | # of steps | # of Big CD |
|---|---|---|---|---|---|
| Sharpen-Trim | ST | $\tanh \Delta^2 \left( \frac{l_j \cosh \Delta^2}{2} \right)^2$ | $\frac{2\pi}{\epsilon_j}$ | 4 | 4 |
| small-Big-small | sBs | $\tanh \Delta^2 \left( \frac{l_j \cosh \Delta^2}{2} \right)^2$ | $2 \times \frac{2\pi}{\epsilon_j}$ | 2 | 2 |
| Big-small-Big | BsB | $2 \times \tanh \Delta^2 \left( \frac{l_j \cosh \Delta^2}{2} \right)^2$ | $\frac{1}{2} \times \frac{2\pi}{\epsilon_j}$ | 2 | 4 |

TABLE S1. Comparison of the different protocols. For all protocols we have defined $\epsilon_j = \sinh \Delta^2 4\pi/l_j$. The number of steps and big controlled-displacements (CD) are given for one full round of the protocol, with circuits for both stabilizers. For the ST and BsB protocols, the number of large CD can be reduced to 2 using qubit measurement and feedback, see Fig. S5.

## D. Qutrit model

In the derivation above, the harmonic bath modes were replaced with a two-level system. It is interesting to note that a similar derivation can be done with a three-level system, yielding slightly improved results. Instead of replacing the bath operators with Pauli matrices $\hat{b}_N \rightarrow (\hat{\sigma}_{x,n} + i\hat{\sigma}_{y,n})/\sqrt{2}$, we can use $\hat{b}_N \rightarrow (\hat{S}_{x,n} + i\hat{S}_{y,n})/\sqrt{2}$, where $\hat{S}_\alpha$ are the spin-1 matrices, $\alpha \in \{x, y, z\}$,

$$\hat{S}_x = \frac{1}{\sqrt{2}} \begin{pmatrix} 0 & 1 & 0 \\ 1 & 0 & 1 \\ 0 & 1 & 0 \end{pmatrix}, \hat{S}_y = \frac{i}{\sqrt{2}} \begin{pmatrix} 0 & -1 & 0 \\ 1 & 0 & -1 \\ 0 & 1 & 0 \end{pmatrix}, \hat{S}_z = \frac{1}{\sqrt{2}} \begin{pmatrix} 1 & 0 & 0 \\ 0 & 0 & 0 \\ 0 & 0 & -1 \end{pmatrix}. \tag{S71}$$

We then follow the same procedure of a Trotter decomposition followed by the replacement of the modular quadrature $\hat{x}_{j,[m_j]} \rightarrow \hat{x}_j$, and we find three new protocols

$$\hat{U}_{\text{qutrit}}^{(ST)} = \begin{cases} e^{-i\frac{\epsilon_{j,q}}{2}\hat{x}_{j,\perp}\hat{S}_y} e^{-il_j c_\Delta \hat{x}_j \hat{S}_x}, & (i) \\ e^{-il_j c_\Delta \hat{x}_j \hat{S}_x} e^{-i\frac{\epsilon_{j,q}}{2}\hat{x}_{j,\perp}\hat{S}_y}, & (ii) \end{cases} \tag{S72a}$$

$$\hat{U}_{\text{qutrit}}^{(BsB)} = e^{-il_j c_\Delta \hat{x}_j \hat{S}_x} e^{-i\epsilon_{j,q}\hat{x}_{j,\perp}\hat{S}_y} e^{-il c_\Delta \hat{x}_j \hat{S}_x}, \tag{S72b}$$

$$\hat{U}_{\text{qutrit}}^{(sBs)} = e^{\frac{-i\epsilon_{j,q}\hat{x}_{j,\perp}\hat{S}_y}{4}} e^{-il c_\Delta \hat{x}_j \hat{S}_x} e^{\frac{-i\epsilon_{j,q}\hat{x}_{j,\perp}\hat{S}_y}{4}}, \tag{S72c}$$

where we have defined

$$\Gamma_{j,\text{qutrit}} \delta t = \tanh \Delta^2 \left( l_j \cosh \Delta^2 \right)^2, \tag{S73}$$

and

$$\epsilon_{j,q} = 2l_j \sinh \Delta^2. \tag{S74}$$

Note that there is no $\theta$ such that $\exp\{i\theta\hat{S}_\alpha\} = -\mathbb{I}$. As a result, instead of choosing the condition Eq. (S59) for the individual pieces of the qutrit unitaries, we choose $\Gamma_{j,\mathrm{qutrit}}\delta t$ such that

$$e^{im_j\hat{x}_{j,\perp}}\hat{U}_{\mathrm{qutrit}}e^{-im_j\hat{x}_{j,\perp}} = \hat{U}_{\mathrm{qutrit}}. \tag{S75}$$

The results comparing the qutrit protocols to the qubit protocols are shown in Fig. S12, where the noise channel is single excitation loss in the oscillator.

## V. FINITE-ENERGY OPERATORS MEASUREMENTS

In the previous section, we analyzed the stabilization circuits in terms of a bath engineering picture. However, we note that one can take a different point of view by replacing the last controlled displacements in Fig. S6 by a qubit measurement followed classical feedback using the rule in Fig. S5. In that situation, we can analyze the information the measured qubit state reveals about the oscillator state.

### A. Stabilizers

In the ST protocol, the feedback version of the sharpening step $(i)$ is equivalent to a one-bit phase estimation protocol [8]. The trimming step $(ii)$, on the other hand, is equivalent to a weak homodyne measurement, more precisely a one-bit measurement of the $\hat{x}_{j,\perp}$ quadrature. Indeed, the Kraus operators associated with the different steps are [7]

$$\hat{K}_{(i),g/e}^{(ST)} = \sqrt{\frac{1 \mp \sin(l_j\cosh\Delta^2\hat{x}_j)}{2}}, \tag{S76a}$$

$$\hat{K}_{(ii),g/e}^{(ST)} = \sqrt{\frac{1 \pm \sin(\epsilon_j\hat{x}_{j,\perp})}{2}}, \tag{S76b}$$

where we have defined $\hat{K}_{g/e} = \langle g/e|\hat{U}(|g\rangle + |e\rangle)/\sqrt{2}$ and we associated $g \leftrightarrow +$, $e \leftrightarrow -$ inside the operator. The probability of measuring $\phi \in \{g, e\}$ when the oscillator is initially in a state $\rho$ is given by

$$P(\phi) = \mathrm{Tr}\left[\hat{K}_\phi^\dagger\hat{K}_\phi\rho\right]. \tag{S77}$$

From $\hat{K}_{(i),g/e}^{(ST)}$, we conclude that the probability of measuring $g$ or $e$ depends on $\sin(l_j\cosh\Delta^2\hat{x}_j) = \mathrm{Im}[\exp\{il_j\cosh\Delta^2\hat{x}_j\}]$, which is a one-bit estimation of the imaginary part of $\hat{T}_{j,\Delta}$. If the displacements in the $\hat{x}_{j,\perp}$ are small, then we can approximate $\sin(\epsilon_j\hat{x}_{j,\perp}) \approx \epsilon_j\hat{x}_{j,\perp}$ and the trimming step can be understood as a one-bit homodyne measurement of the quadrature $\hat{x}_{j,\perp}$ with a precision set by $\epsilon_j$. We can therefore understand the ST protocol as splitting in two the measurement of the finite-energy operator $\hat{d}_{j,\Delta} \propto \hat{x}_{j,[m_j]} + i\hat{x}_{j,\perp}\tanh\Delta^2$, with the sharpening step measuring $\hat{x}_{j,[m_j]}$ and the trimming step measuring $\hat{x}_{j,\perp}$. The full measurement of $\hat{d}_{j,\Delta}$ is then approximated by performing these two measurements one after the other.

In contrast, the sBs and BsB protocol aim to measure the finite-energy stabilizers in a single shot. More precisely, the Kraus operator for the sBs protocol is given by (up to an irrelevant global phase)

$$\hat{K}_{g/e}^{(sBs)} = \frac{1}{2}\left[e^{\frac{il_j\cosh\Delta^2\epsilon_j}{8} - \frac{i\pi}{4}}\sqrt{1 \mp \sin(l_j\cosh\Delta^2\hat{x}_j - \epsilon_j\hat{x}_{j,\perp}/2)} \right.$$
$$\left. + e^{-\frac{il_j\cosh\Delta^2\epsilon_j}{8} + \frac{i\pi}{4}}\sqrt{1 \mp \sin(l_j\cosh\Delta^2\hat{x}_j + \epsilon_j\hat{x}_{j,\perp}/2)}\right]. \tag{S78}$$

Taking each term in the above sum independantly, they can be understood as performing a phase estimation of two slightly rotated lattices in the directions $\hat{x}_j \pm \delta_\epsilon\hat{x}_{j,\perp}$, with $\delta_\epsilon$ a small angle set by $\epsilon$. Measuring the *superposition* of these two lattices leads to a Moiré pattern, with the appearance of a larger superlattice of size $\propto 1/\epsilon$. As long as the state is confined within the central unit cell of that Moiré superlattice, the wavefunction of the stabilized GKP approximates the desired Gaussian envelope. We can also compute the

Kraus operators for the BsB protocol,

$$\hat{K}_{g/e}^{(BsB)} = \frac{1}{2} \left[ e^{\frac{il_j \cosh \Delta^2 \epsilon_j}{2} - i\frac{\pi}{4}} \sqrt{1 \pm \sin(l_j \cosh \Delta^2 \hat{x}_j + 2\epsilon_j \hat{x}_{j,\perp})} \right.$$
$$\left. + e^{\frac{-il_j \cosh \Delta^2 \epsilon_j}{2} + i\frac{\pi}{4}} \sqrt{1 \pm \sin(l_j \cosh \Delta^2 \hat{x}_j - 2\epsilon_j \hat{x}_{j,\perp})} \right]. \tag{S79}$$

## B. Pauli Operators

The measurement point of view becomes particularly useful when it comes to the measurement of logical Pauli operators. In particular, one can improve on the phase estimation method used in Refs. [7, 9] by considering finite-energy Pauli operators instead of the ideal ones.

In order to measure the Pauli operator, we aim to perform the equivalent of the qubit-qubit circuit illustrated in Fig. S7a, where the controlled-NOT (CNOT) operation is described by

$$\hat{U}_{\text{CNOT}} = e^{-i\frac{\pi}{2}\hat{\Pi}_e \otimes \sigma_x}, \tag{S80}$$

where $\hat{\Pi}_e$ is the projector onto the excited state of the control qubit. It is easy to measure other Pauli operators by replacing $\hat{\Pi}_e \to \hat{\Pi}_-$, where $\hat{\Pi}_-$ is the projector onto the -1 eigenstate of the Pauli operator to be measured.

Denoting the ideal GKP logical Pauli operators $\hat{P}_0 = \exp\{il_j \hat{x}_j/2\}$, we define the analogue of $\hat{d}_{j,\Delta}$, see Eq. (S40), for the finite-energy Pauli $\hat{P}_\Delta$,

$$\hat{f}_{j,\Delta} = \frac{\frac{\hat{x}_{j,[4\pi/l_j \cosh \Delta^2]}}{\sqrt{\tanh \Delta^2}} + i\hat{x}_{j,\perp}\sqrt{\tanh \Delta^2}}{\sqrt{2}}, \tag{S81}$$

where the variable $\hat{x}_j$ is taken modular with a period twice as large as in $\hat{d}_{j,\Delta}$. We want to distinguish between the $\pm 1$ eigenstates of $\hat{P}_\Delta$, which are also eigenstates of this new squeezed-annihilation-like operator

$$\hat{P}_\Delta |\psi\rangle = |\psi\rangle \Leftrightarrow \hat{f}_{j,\Delta} |\psi\rangle = 0, \tag{S82a}$$

$$\hat{P}_\Delta |\psi\rangle = -|\psi\rangle \Leftrightarrow \hat{f}_{j,\Delta} |\psi\rangle = f|\psi\rangle, \tag{S82b}$$

where we have defined $f = \pi/l_j \sqrt{2/(\cosh \Delta^2 \sinh \Delta^2)}$. From the above relation, we see that, in the logical subspace, $\hat{f}_{j,\Delta}/f$ acts as a projector onto the -1 eigenstate of $\hat{P}_\Delta$. However, $\hat{f}_{j,\Delta}$ is not a Hermitian operator and the immediate generalization of Eq. (S80), $\exp\{-i\pi/2(\hat{f}_{j,\Delta}/f)\hat{\sigma}_x\}$, is not a unitary. In order to turn this into a unitary, one option is to approximate $\hat{f}_{j,\Delta}/f \to \text{Re}[\hat{f}_{j,\Delta}/f] = (\hat{f}_{j,\Delta} + \hat{f}_{j,\Delta}^\dagger)/2f$, leading to

$$\hat{U}_{\text{Phase Est}} = e^{-i\frac{\pi}{2}\frac{(\hat{f}_{j,\Delta} + \hat{f}_{j,\Delta}^\dagger)}{2f}\hat{\sigma}_x},$$
$$= e^{i\frac{\pi}{2f\sqrt{2\tanh \Delta^2}}\hat{x}_{j,[4\pi/l_j \cosh \Delta^2]}\hat{\sigma}_x}, \tag{S83}$$
$$= e^{i\frac{l_j \cosh \Delta^2}{4}\hat{x}_j \hat{\sigma}_x}.$$

Approximating $\cosh \Delta^2 \approx 1$, this circuit corresponds to the one-bit phase estimation protocol [8] illustrated in Fig. S7b which was realized in Refs. [7, 9]. In the limit of large GKP states, $\Delta \to 0$, we have $\hat{f}_{j,0} = \hat{f}_{j,0}^\dagger$ and the circuit becomes an exact CNOT. However, since $\hat{f}_{j,\Delta}|\psi\rangle = f|\psi\rangle$ does not imply that $\hat{f}_{j,\Delta}^\dagger|\psi\rangle \neq f^*|\psi\rangle$, the unitary above is not exactly the desired CNOT gate and induces measurement errors as shown in Fig. S8.

Instead of approximating the CNOT gate using the unitary Eq. (S83), consider the evolution generated by a Jaynes-Cummings-like coupling

$$\hat{U}_{\text{JC}} = e^{-i\frac{\pi}{2}\frac{(\hat{f}_{j,\Delta}\hat{\sigma}_+ + \hat{f}_{j,\Delta}^\dagger \hat{\sigma}_-)}{f}},$$
$$= e^{\frac{-il_j \cosh \Delta^2}{4}\left(\hat{x}_{j,[2m_j]}\hat{\sigma}_x + \hat{x}_{j,\perp}\hat{\sigma}_y \tanh \Delta^2\right)}, \tag{S84}$$

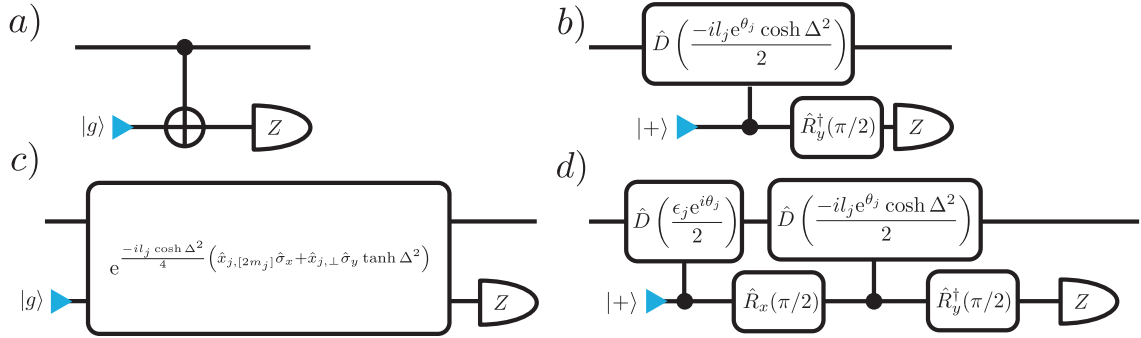

FIG. S7. Circuits for measurement in the logical subspace. a) Ideal measurement circuit for the Pauli Z operator. b) Phase estimation protocol. c) Circuit corresponding to the unitary in Eq. (S84). d) Approximate version of the circuit in c), leading to better results than the phase estimation protocol.

where we have defined $\hat{\sigma}_\pm = (\hat{\sigma}_x \mp i\hat{\sigma}_y)/2$. In the limit $\Delta \to 0$, this unitary is equivalent to the phase estimation unitary, Eq. (S83), and to the desired CNOT, Eq. (S80). As in the stabilization circuit, this unitary seems challenging to implement directly and we look for simplifications using Trotter decompositions. From the four possible decompositions of first and second order described in Sect. IV C, we find that the only one leading to improved results over the phase-estimation protocol Eq. (S83) is

$$\hat{U}_{\mathrm{JC},(ii)} = \mathrm{e}^{\frac{-il_j \cosh \Delta^2}{4} \hat{x}_j \hat{\sigma}_x} \mathrm{e}^{-i\frac{\epsilon_j}{4} \hat{x}_{j,\perp} \hat{\sigma}_y},$$
$$\approx \hat{U}_{\mathrm{JC}},$$

(S85)

where we used the same definition for $\epsilon_j$ as before, see Eq. (S60). We used the index $(ii)$ to make the connection to the trimming circuit used in Sect. IV C. The circuit equivalent of this unitary, expressed using controlled displacements, is illustrated in Fig. S7d. Both $\hat{U}_{\mathrm{JC},(ii)}$ and $\hat{U}_{\mathrm{Phase\ Est}}$ leave the oscillator displaced by $l_j/4$, which can be corrected by a displacement after the measurement. As shown in Fig. S8, $\hat{U}_{\mathrm{JC},(ii)}$ leads to a higher measurement fidelity than the phase estimation protocol. We note that a similar circuit was independantly discovered in Ref. [10].

Finally, we have presented some strategies for the measurement of finite-energy Paulis using qubits, but robust measurements can also be made using homodyne measurements, as was introduced in the original paper by Gottesman, Kitaev and Preskill [2].

## VI. FINITE-ENERGY GATES

As shown in Sect. III C, the logical Pauli operators are not unitary operators. One consequence is that applying finite-energy Pauli gates is not a unitary operation and necessitate the use of an ancilla. Fortunately, we showed earlier that one stabilization step of the protocols sBs and ST applies a logical Pauli operation, with the sBs being closer to the finite-energy Pauli than either of the ST steps. Consequently, the individual steps of the ST and sBs stabilization protocols can double as logical gates. While using the stabilization protocol for logical gates allow a better control of the GKP envelope than simple displacements, we note that they take longer to execute and expose the oscillator to ancilla errors. As mentioned in the main Letter, the impact of ancilla errors could be mitigated by making use of a biased-noise ancilla where bit-flip are suppressed. We leave the in-depth study of logical gate strategies for future work.

Since the envelope operator is defined as a function of $\hat{n}$, gates based on a lattice rotation $\exp\{i\theta\hat{n}\}$ are identical to their ideal counterpart. For example, for the square lattice, the Hadamard gate corresponds to a quarter-turn rotation of phase space, $\exp\{i\pi\hat{n}/2\}$, and this operation is the same for finite-energy and ideal GKP states. In practice, the orientation of phase space is defined with respect to a phase reference, and this kind of rotation can be done instantaneously in software by changing the phase reference of all subsequent pulses.

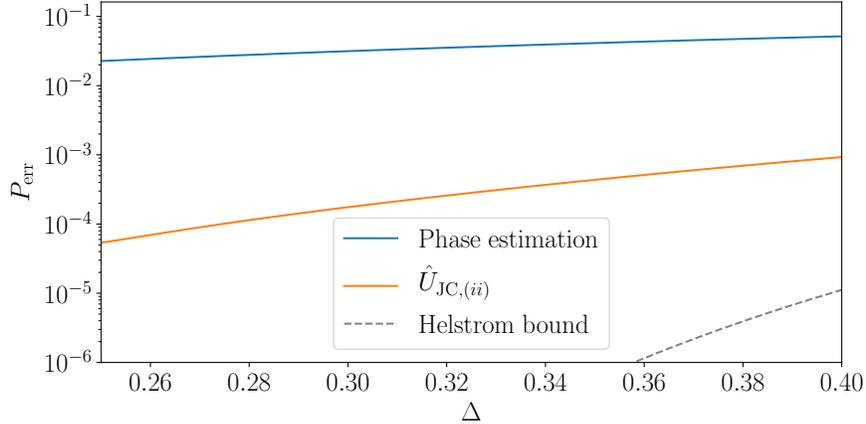

FIG. S8. Measurement error probability as a function of the GKP state size, $\Delta$. We take a square lattice GKP and compute the average probability of a measurement error $P_{\text{err}} = (P(g|1_\Delta) + P(e|0_\Delta))/2$. The blue curve corresponds to the phase estimation protocol and the orange curve to $\hat{U}_{JC,(ii)}$. Due to the finite overlap between the codewords, the measurement error is bounded from below by the Helstrom bound, $P_{\text{err}} \geq (1 - \sqrt{1 - |\langle 0_\Delta | 1_\Delta \rangle|^2})/2$ (grey dashed line).

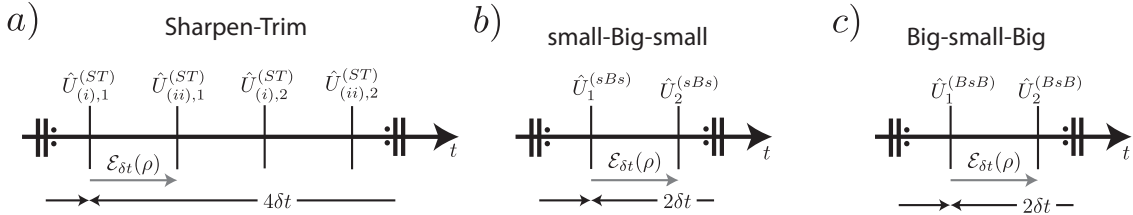

FIG. S9. One full round for the ST (a), sBs (b) and BsB (c) protocols. One full round lasts for $4\delta t$ for the ST protocol, while one round lasts for $2\delta t$ for the sBs and BsB protocols. The illustrated rounds are repeated enough times to extract the logical information encoded in the GKP states.

## VII. SIMULATIONS

In this section, we provide more information about the simulation results presented in the main Letter.

### A. Oscillator Errors

In order to focus on the effect of oscillator errors, we first consider perfect and instantaneous oscillator-ancilla circuits. In that situation, the effect of the oscillator-ancilla circuits can be computed through a Kraus map $\{\hat{K}_g, \hat{K}_e\}$, with Kraus operators

$$\hat{K}_{g/e,j} = \langle g/e | \hat{U}_j^{(\alpha)} | + \rangle. \tag{S86}$$

The exact expressions for these Kraus operators are easily computed by adding a displacement to the equations in Sect. V A. We then use the Kraus operators to compute the unconditional evolution of the oscillator density matrix,

$$\rho_{n+1} = \hat{K}_{g,j} \rho_n \hat{K}_{g,j}^\dagger + \hat{K}_{e,j} \rho_n \hat{K}_{e,j}^\dagger. \tag{S87}$$

Between the applications of the stabilization unitaries, we let the oscillator evolve under some noise channel $\mathcal{E}(\rho)$, with the full sequence for one round of each protocol schematically illustrated in Fig. S9.

In order to extract the channel fidelity, we first prepare each of the Pauli eigenstates. To do so, we initially take the oscillator to be in a vacuum state $|0\rangle$, and then stabilize the GKP manifold of the desired size by

repeatedly applying the sequences in Fig. S9. The stabilization protocols do not distinguish between the logical states $|\mu_\Delta\rangle$ and because the vacuum state has non-zero overlap with both logical states, $\langle 0|\mu_\Delta\rangle \neq 0$, the oscillator ends up in a mixed logical state. Moreover, we also consider oscillator errors during state preparation, which induces more decoherence inside the logical manifold. After having prepared the logical manifold, we use the circuit in Fig. S7b to measure the logical Pauli operators, which projects the GKP state onto one of the Pauli eigenstates. Since the ideal circuits do not ideally distinguish between the finite-energy logical states, we apply the circuits twice, and post-select on measuring the same result twice in a row.

Instead of doing the state preparation procedure above, one option would have been to directly input the finite-energy states $|\mu_\Delta\rangle$ into the simulation. However, we note that, although this is our goal, the steady state of the stabilization circuits is not *exactly* the finite-energy GKP manifold. Indeed, some approximations were made while deriving the stabilization circuits, namely the Trotter decomposition and the "qubitization." Moreover, because of a finite ratio between error rate and error correction rate $\Gamma$, the stabilization steady states contain a mixture of correctable errors. Starting with a pure logical state $|\mu_\Delta\rangle$, there is a fast timescale $t_e$ associated with the equilibration of small errors and a (hopefully) long timescale $t_L$ associated with the lifetime of logical information. For small error channels, we have $t_e \ll t_L$, and the two timescales can be easily distinguished. However, for stronger error channels near the break-even point we have $t_e \lesssim t_L$, and the two effects are harder to distinguish. Since we aim to extract the logical lifetime, we have chosen a state preparation procedure that already contains the equilibration effects such that $t_e \to 0$. Finally, the state preparation method we used is also closer to the procedure used experimentally, which makes the connection between theory and experiment easier.

For each of the prepared Pauli eigenstates, we apply the stabilization sequences of Fig. S9 and compute the evolution of the associated logical Pauli expection value. The noise channels we consider in Fig.3 of the main Letter are an amplitude damping channel described by

$$\dot{\rho} = \kappa \mathcal{D}[\hat{a}]\rho, \tag{S88}$$

and a white noise dephasing channel described by

$$\dot{\rho} = \kappa_\phi \mathcal{D}[\hat{a}^\dagger \hat{a}]\rho. \tag{S89}$$

Examples of a time evolution for the $+Z$ state are shown in Fig. S10c and Fig. S11c. The former figure is for a perfect oscillator, while the latter is computed from $\kappa \delta t = 0.0038$. Starting from the time of the state projection, we fit a linear function over a short time window in order to extract the decay constant of the logical information and obtain six decay rates $\{\gamma_k\}$, one for each Pauli eigenstate $k \in \{\pm x, \pm y, \pm z\}$. For the ST and sBs protocols, we account for the fact that the stabilization steps apply a logical Pauli gate. In order to simplify the calculation of the channel fidelity, we take a Pauli error model for the logical GKP manifold, $\mathcal{E} = \{\sqrt{p_I}\mathbb{1}, \sqrt{p_x}\hat{\sigma}_x, \sqrt{p_y}\hat{\sigma}_y, \sqrt{p_z}\hat{\sigma}_z\}$. Since X eigenstates are affected by both Y and Z errors, we have $\gamma_{\pm x}\delta t \approx p_y + p_z$, and similarly for the other decay rates. The remaining parameter, $p_I$, is set by the normalization condition $p_I = 1 - p_x - p_y - p_z$. Setting $\gamma_\alpha = (\gamma_{+\alpha} + \gamma_{-\alpha})/2$ with $\alpha \in \{x, y, z\}$, we get

$$\begin{pmatrix} p_x \\ p_y \\ p_z \end{pmatrix} = \frac{1}{2} \begin{pmatrix} -1 & 1 & 1 \\ 1 & -1 & 1 \\ 1 & 1 & -1 \end{pmatrix} \begin{pmatrix} \gamma_x \delta t \\ \gamma_y \delta t \\ \gamma_z \delta t \end{pmatrix} \tag{S90}$$

The effect of the logical qubit channel $\mathcal{E}$ can be put into matrix form using

$$\mathcal{E}_{ij} = \frac{\text{Tr}(\hat{\sigma}_i \mathcal{E}(\hat{\sigma}_j))}{2}, \tag{S91}$$

and the fidelity of the logical channel is then given by

$$\mathcal{F} = \frac{\text{Tr}(\mathcal{E})}{4}. \tag{S92}$$

In terms of the fitted decay rates, we get

$$\mathcal{F}_{\text{GKP}} = 1 - \frac{\sum_{k \in \{\pm x, \pm y, \pm z\}} \gamma_k \delta t}{2}. \tag{S93}$$

We remark that the above formula is only valid at short times $\gamma_k \delta t \ll 1$ when the loss of logical information is well described by a linear behavior.

For the Fock encoding and single-excitation loss channel, the fidelity can be computed exactly,

$$\mathcal{F}_{\text{Fock},\kappa} = \frac{1 + 2e^{\frac{-\kappa\delta t}{2}} + e^{-\kappa\delta t}}{4},$$

$$\approx 1 - \frac{\kappa\delta t}{2}. \tag{S94}$$

We can also compute the fidelity exactly for the dephasing channel,

$$\mathcal{F}_{\text{Fock},\kappa_\phi} = \frac{1 + e^{-\kappa_\phi\delta t}}{2},$$

$$\approx 1 - \frac{\kappa_\phi\delta t}{2}. \tag{S95}$$

### 1. Uncorrected GKP

In order to better appreciate the effect of the stabilization protocol, we also compute the logical channel fidelity without any stabilization, with the results illustrated by the dashed lines in Figure 3 of the main Letter. To compute these curves, we used the same state preparation and fitting procedure, but after the Pauli measurement we let the oscillator evolve freely without applying the stabilization circuits. For the single excitation loss channel, we can roughly estimate the fidelity of the uncorrected GKP states. Taking the loss channel as a non-Hermitian Hamiltonian $\hat{H} = -i\kappa\hat{n}/2$, we get the evolution of the logical $\hat{X}_0 = \exp\{-il_1\hat{p}\}$ operator

$$\langle \dot{\hat{X}}_0 \rangle = -i\langle [\hat{X}_0, \hat{H}]\rangle,$$

$$= -\frac{\kappa}{4}\left\langle \hat{X}_0 \left( l_1\hat{x} + \frac{l_1^2}{4}\right)\right\rangle \tag{S96}$$

Taking the approximation that $\langle \hat{X}_0\hat{x}\rangle \approx \langle \hat{X}_0\rangle\langle \hat{x}\rangle = 0$, we get

$$\langle \hat{X}_0\rangle(t) \approx e^{-\frac{\kappa t l_1^2}{16}}\langle \hat{X}_0\rangle(0). \tag{S97}$$

Performing a similar computation for the $\hat{Z}_0$ and $\hat{Y}_0$ operators, we get

$$\mathcal{F}_{\text{uncorrectedGKP}} = \frac{1 + e^{\frac{-\kappa\delta t l_1^2}{16}} + e^{\frac{-\kappa\delta t l_2^2}{16}} + e^{\frac{-\kappa\delta t|l_1 e^{i\theta_1} + l_2 e^{i\theta_2}|^2}{16}}}{4},$$

$$\mathcal{F}_{\text{uncorrectedGKP}} \approx 1 - \frac{\kappa\delta t}{64}\left(l_1^2 + l_2^2 + |l_1 e^{i\theta_1} + l_2 e^{i\theta_2}|^2\right). \tag{S98}$$

For the square lattice, this expression simplifies to $\mathcal{F}_\square \approx 1 - \kappa\delta t l_\square^2/16 = 1 - \kappa\delta t\pi/4$. Note that the above fidelity does not depend on the GKP size, which is consistent with the dashed lines in Fig. 3. Comparing the infidelity of the uncorrected GKP against the Fock encoding, we get

$$\frac{1 - \mathcal{F}_{\text{uncorrectedGKP}}}{1 - \mathcal{F}_{\text{Fock},\kappa}} = \frac{1}{32}\left(l_1^2 + l_2^2 + |l_1 e^{i\theta_1} + l_2 e^{i\theta_2}|^2\right) \geq \frac{\pi}{2}. \tag{S99}$$

The lower bound of $\pi/2$ is obtained for the square GKP lattice. Since this factor does not depend on $\delta t$ or $\kappa$, it (unsurprisingly) indicates that one cannot do better than the Fock encoding without error correction. The fact that this ratio is larger than 1 also reflects the fact that GKP states contain more excitations than the Fock encoding, and the consequent higher rate of excitation loss imposes an overhead that the QEC must surmount in order to surpass the lifetime of the Fock encoding.

### 2. Tunneling between sites of the superlattice

One potential issue with the stabilization protocol is the presence of the superlattice. Indeed, only the central site of the superlattice minimizes the excitation number, and a population in the other sites is more

susceptible to non-linear error channels such as an oscillator self-Kerr term or oscillator dephasing. Because of its smaller superlattice and larger displacements in a single step, the BsB protocol is more prone to tunneling errors, even without oscillator decoherence. This explains why errors saturate at a higher level for the BsB protocol compared with the sBs and ST protocols.

Counter-intuitively, tunneling can also be caused by excitation loss. Indeed, *conditioned* on having lost an excitation, a state can be left with more excitations than it started with. The state $(|0\rangle + |4\rangle)/\sqrt{2}$ provides a simple example of this phenomenon, where the average excitation number increases from 2 to 3 after having lost an excitation. The GKP states are similar in that they contain a superposition of many Fock states. After having lost an excitation, GKP states contain more energy, and therefore extend further in phase space, which in turn leads to an increase probability of tunneling. In simulations, we observed that tunneling can become non-negligible for some values of $\kappa\delta t$, which manifests in part by a steady increase in average excitation number during the stabilization. The previous discussion superficially seems to contradict the intuition that, on average, amplitude damping reduces the amount of energy contained in the oscillator. This is indeed the case when we take into account the evolution of the oscillator conditioned on *not* losing an excitation ("no-jump" evolution). However, we remark that energy in injected in the oscillator at each stabilization step, and that the ratio of energy injected against energy lost is very large for error rates small enough to protect the logical information, $\kappa\delta t \ll 1$. Consequently, the energy confinement induced by amplitude damping is not sufficient, and we rely instead on the interference effects of the stabilization protocol.

We also remark that only the BsB protocol leads to a poor confinement for the hexagonal GKP code. This is because $l_{\bigcirc} > l_{\square}$, and the controlled displacements in a single step lead to a higher tunneling probability. In practice, we suspect that some combination of BsB and sBs, for example alternating between the 2 protocols, will be optimal for QEC. This would allow to combine the larger error correction rate of the BsB protocol with the better confinement of the sBs protocol.

Finally, we note that tunneling increases for the qutrit model compared with the qubit model. This is because individual controlled-displacements in the qutrit model displace the oscillator by $\{0, \pm l\}$, compared with the qubit model where the displacements are given by $\pm l/2$. Because of these larger displacements, the probability of tunneling increases. Figure S12 shows that for all protocols, the error saturation is higher for the qutrit protocols. However, the larger correction rate of the qutrit protocol leads to an increased break-even point. For the qutrit version of the BsB protocol, the confinement is very poor, and the middle bend in the curve is due to the finite Hilbert space used in the simulations. While it is possible to generalize our stabilization protocols to larger $d-$level ancillas, we expect the confinement to become less effective in this situation.

### 3. More comments on Fig. 3

For standard qubit QEC codes such as a surface code with order $L^2$ qubits, the probability of a logical error can be estimated using

$$p_L \sim \left(\frac{p}{p_c}\right)^{\frac{d(L)-1}{2}}, \tag{S100}$$

where $d(L) \sim L$ is the distance of the code, $p$ is the probability of a physical error in single qubit and $p_c$ is the threshold. When tracing the logical error rate $p_L$ against the physical error rate $p$ on a logarithmic scale, $d$ appears in the slope of the curve. If the physical error probability is below threshold, $p < p_c$, the logical error probability can be reduced arbitrarily by increasing the code size $L$.

In contrast, for our GKP stabilization scheme, an estimate of the logical error rate below the break-even point is given by

$$p_L \sim \max\left[p_{\text{sat}}(\Delta), \left(\frac{p}{p_{\text{be}}(\Delta)}\right)^{\frac{d(\Delta)-1}{2}}\right], \tag{S101}$$

where we have defined a saturation error $p_{\text{sat}}(\Delta)$ and a break-even point $p_{\text{be}}(\Delta)$ which both depend on $\Delta$. In Fig. 3, the break-even point $p_{\text{be}}(\Delta)$ is given by the crossing point between a fidelity curve and the Fock encoding curve (red). Because $p_{\text{be}}(\Delta)$ depends on the GKP size, it is not a threshold and we use the term break-even instead. The saturation error $p_{\text{sat}}(\Delta)$ is given by the value of the channel fidelity when

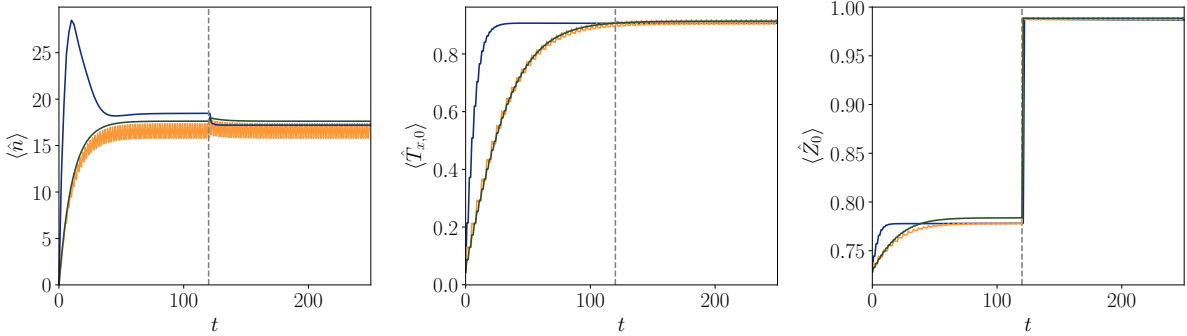

FIG. S10. Time evolution under the stabilization cicuits for the square GKP without oscillator decoherence and with $\epsilon = 0.1$. The oscillator is initialized in vacuum and projected onto a +Z eigenstate at $t = 120\delta t$ (grey dashed line). Blue lines correspond to the BsB protocol, dark green lines to the sBs protocol and orange lines to the ST protocol. a) Average number of excitations as a function of time. In steady state, the sBs and BsB protocols keep $\langle \hat{n} \rangle$ constant. In contrast, for the ST protocol, the excitation number increases during the sharpening steps $(i)$ and then decreases during the trimming steps $(ii)$. Out of the three protocols, only the BsB exhibits the same excitation number overshoots as in the continuous model Fig. S3a. b) Expectation value of the ideal stabilizer as a function of time. The three protocols stabilize at roughly the same value, indicating that the states generated by the different protocols are of the same quality. We attribute the slight steady state differences in excitation number and $\langle \hat{T}_{x,0} \rangle$ between the protocols to the different approximations that were made while deriving them. Since the stabilization rate $(\Gamma)$ is the same for the sBs and ST protocol, they converge towards their steady-state value at the same rate, while the BsB converges roughly twice as fast, see Tab. S1. For the ST protocol, $\langle \hat{T}_{x,0} \rangle$ evolves on a cycle of $4\delta t$. The first two time steps are an increase (sharpening) followed by a decrease (trimming). Then, its value stays constant for two times steps during the stabilization of the +1 eigenstate of $\hat{T}_p$. c) Expectation value for $\langle \hat{Z}_0 \rangle$ as a function of time. For all protocols, the expectation value increases sharply at $t = 120\delta t$ due to the projective measurement. Since there are no errors in the oscillator, the lifetime of the logical information for all protocols is very long and cannot be resolved on this timescale.

$\kappa, \kappa_\phi \to 0$. We note that it is difficult to apply the traditional notion of distance to the GKP case. Indeed, the eigenvalues of the GKP stabilizers are continuous, and natural error models are also continuous. Moreover, the codewords are not exactly orthogonal, and the finite-energy GKP code cannot formally correct against *any* error channel, even the identity channel. To circumvent these difficulties, we define the distance $d(\Delta)$ as slope of the fidelity curve just below the break-even point, in analogy with Eq. (S100).

Even with a decoherence-free oscillator $(p = 0)$, the logical error rate saturates at $p_{sat}(\Delta)$ for a given GKP size $\Delta$. This is due in part to the finite overlap between codewords $\langle 0_\Delta | 1_\Delta \rangle$ and in part due to tunneling between sites of the superlattice, see the previous discussion. These errors disappear as the GKP size increases, $p_{sat}(0) \to 0$. However, for a finite physical error probability, $p \approx \kappa \delta t$, the second term in the max function cannot be reduced arbitrarily. Indeed, increasing the GKP size leads to an increase in the distance $d(\Delta)$, but a decrease in the break-even point, $p_{be}(\Delta)$. This is because the break-even point depends on the ratio between the error rate and the correction rate, $\kappa/\Gamma$, and the correction rate decreases with larger GKP states, $\Gamma \propto \Delta^2$. In short, for a fixed physical error rate, the GKP size needs to be optimized to balance the number of errors that can be corrected, $d(\Delta)$, against the rate at which these errors are corrected, $\Gamma$.

## B. Ancilla errors

In order to study the effect of ancilla decoherence on the GKP logical manifold, we take stabilization protocols where each step takes a finite time $\delta t$. We consider that the controlled displacements are generated by time evolution under the Hamiltonian

$$\hat{H}_{CD} = g_z(\hat{a}e^{i\varphi} + \hat{a}^\dagger e^{-i\varphi})\hat{\sigma}_z, \tag{S102}$$

and that the qubit rotations are instantaneous and perfect. We also consider that qubit reset is instantaneous and perfect. The unitary evolution generated by Eq. (S102) is given by

$$\hat{U}_{CD} = e^{-it_{CD}\hat{H}_{CD}}, \tag{S103}$$

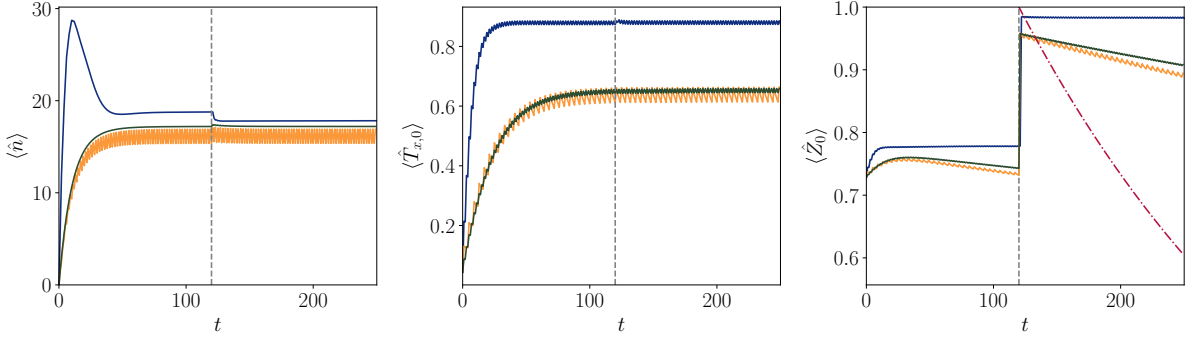

FIG. S11. Similar time evolution as in Fig. S10, but with an excitation loss channel, $\kappa\delta t = 0.0038$. a) The excitation number as a function of time is similar to the no-loss case for all protocols. For the BsB protocol, the average excitation number is slighly larger due to tunneling between sites of the superlattice. b) Because of the finite ratio $\kappa/\Gamma$, the QEC steps do not correct all errors and the steady state of all protocols consists of some mixture of displacement errors on top of the coherent errors induced by the envelope $\hat{E}_\Delta$. The BsB has a correction rate twice as large as the other protocols, and its steady state is of better quality. Since the sBs and ST protocol have the same error correction rate $\Gamma$, the steady state value of $\langle\hat{T}_{x,0}\rangle$ is reduced by the same amount. The BsB and sBs protocol exhibit a two-step oscillation since displacement errors in $\hat{x}$ are only corrected every two steps. c) The time evolution of the logical Pauli operator shows how the BsB protocol preserves the logical information longer. The red dotted-dashed line is a guide for the eye indicating the $\exp\{-\kappa t\}$ decay of the $|1\rangle$ Fock state.

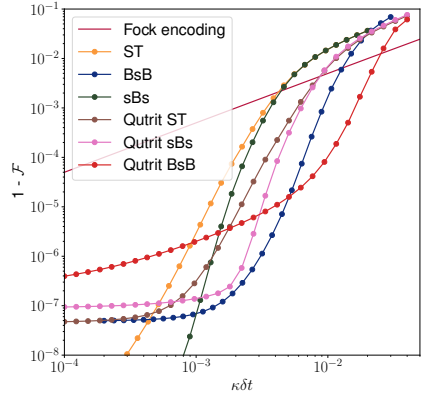

FIG. S12. Channel fidelity with the qutrit version of the stabilization protocols, see Sect. IV D. We chose a square lattice and a single GKP size set by $\epsilon = 0.1$. The orange, blue and dark green curves are the same as in Fig. 3. For all qutrit protocols, the break-even point with the Fock encoding is increased compared to the qutrit version of the protocol. The qutrit protocols exhibit behaviors similar to their qubit counterpart. First, since the QEC rate is the same for the ST and sBs protocols, their break-even point is the same. However, since the sBs protocol is derived from a higher-order approximation of the stabilization unitary, the sBs protocol has a larger slope below the break-even point and better corrects errors. Because the controlled displacements are larger in the qutrit version, the BsB confines the GKP states less than in the qutrit version. As a consequence, tunneling to other sites of the superlattice is worse and limit the fidelity. Simulations show a corresponding increase in excitation number during the stabilization using the qutrit BsB protocol.

and a controlled displacement $C\hat{D}(\beta)$ is obtained by choosing $g_z t_{\mathrm{CD}} e^{-i\varphi} = i\beta/(2\sqrt{2})$.

### 1. Ancilla decay

The most damaging type of physical error for the logical information is ancilla decay (or bit flips) during the controlled displacements (CD). For each CD, we simulate the evolution of the oscillator-qubit system

under the master equation

$$\dot{\rho} = -i[\hat{H}_{\mathrm{CD}}, \rho] + \gamma_1 \mathcal{D}[\hat{\sigma}_-]\rho, \tag{S104}$$

where $\gamma_1 = 1/T_1$ is the decay rate of the ancilla.

In order to gain a better understanding of the effect of ancilla decay, we study the simpler case of random bit flips. If a bit flip happens at exactly $t_{\mathrm{CD}}/2$, then instead of the desired CD we obtain

$$\hat{U} = \mathrm{e}^{-i\frac{t_{\mathrm{CD}}}{2}\hat{H}_{\mathrm{CD}}} \hat{\sigma}_x \mathrm{e}^{-i\frac{t_{\mathrm{CD}}}{2}\hat{H}_{\mathrm{CD}}}, \tag{S105}$$
$$= \mathbb{I}.$$

For large controlled displacements, $C\hat{D}(l)$, this means that instead of applying a logical Pauli operation as intended, the operation is proportional to the identity which is a logical error. More generally, if the bit flip happens at $\tau \in [0, t_{\mathrm{CD}}]$, the resulting unitary is given by $\hat{U} = C\hat{D}[\beta(2\tau/t_{\mathrm{CD}} - 1)]$, which leads to a logical error if $\tau \in [t_{\mathrm{CD}}/4, 3t_{\mathrm{CD}}/4]$. For $\tau \in [0, t_{\mathrm{CD}}/4]$, the resulting displacements differ from the ideal ones by one full lattice constant plus correctable errors, which is a correctable error. When $\gamma_1 t_{\mathrm{CD}} \ll 1$, we can approximate the probability of a decay event during the CD to be $p \approx \gamma_1 t$. Given that the probability of a decay event is uniformly distributed in time during the controlled displacement, we conclude that the probability of a logical error for one CD is given by $p \approx \gamma_1 t/2$.

Displacement errors induced by a decay event during a small controlled displacement, $C\hat{D}(\epsilon)$, are correctable, and we conclude that the fidelity is limited by ancilla decay during the large controlled displacements, $C\hat{D}(l)$. In particular, logical errors due to ancilla decay do not depend on the GKP size.

It is possible to partly mitigate qubit decay errors by replacing the final large CD in the ST and BsB protocols by a qubit measurement followed by classical feedback as in Fig. S5. This can effectively reduce by a factor of 2 the number of large CD and the corresponding logical error rate. However, depending on the architecture, this can make the protocol longer due to the delay between the qubit measurement of the classical feedback.

Finally, the channel fidelity obtained in Fig. 4a represents a situation where the oscillator is constantly entangled with the ancilla, and therefore constantly exposed to ancilla decay. By adding an intentional delay between each stabilization circuit, one can reduce the fraction of the time the oscillator is exposed to ancilla errors, at the cost of a slower correction of oscillator errors. In practice, the repetition rate $(1/\delta t)$ of the stabilization scheme has to be optimized in order to strike a balance between a faster correction of oscillator errors obtained by reducing $\delta t$ and a reduction of ancilla-induced errors obtained by increasing $\delta t$.

### 2. Ancilla dephasing

In contrast to decay, dephasing errors commute with the controlled displacements, $[\hat{\sigma}_z, \hat{U}_{\mathrm{CD}}] = 0$, and, as we show, a single phase error $\hat{\sigma}_z$ cannot induce a logical error in the GKP manifold. To study the effect of dephasing, we consider the evolution of the oscillator-ancilla system under the master equation

$$\dot{\rho} = -i[\hat{H}_{\mathrm{CD}}, \rho] + \frac{\gamma_\phi}{2} \mathcal{D}[\hat{\sigma}_z]\rho. \tag{S106}$$

In Fig. S13, we illustrate how dephasing errors affect the GKP states for the sBs protocol. The analysis for the two other protocols is similar. To simplify the illustration, we choose $l_j e^{i\theta_j} = l$ and $\cosh \Delta^2 \approx 1$. First, a dephasing error during the first small CD (a) leads to the sign of the large CD being reversed. However, this error differs from the errorless situation by one full lattice constant $l$, and is therefore correctable. Second, a dephasing error during the large CD (b) leads to the sign of the last CD to be reversed, but this induces a displacement error of $\epsilon$ which is correctable. Finally, a dephasing error during the last small CD (c) has no effect on the GKP state. Since dephasing errors commute with the controlled displacements, the error locations considered in Fig. S13 are sufficient to get a complete picture.

To further understand the robustness of the GKP code to ancilla dephasing, consider the limit of a classical ancilla that dephases instantly, $\gamma_\phi \to \infty$. In that situation, the stabilization circuits correspond to a channel that induces a random displacement of length $\epsilon$ (modulo $l$) on the oscillator at each step. Therefore, even when $\gamma_\phi \to \infty$, the lifetime of the logical information stays finite. Of course, in such a limit, using the ancilla does not lead to error-correction, and the maximal GKP lifetime is obtained by not performing the stabilization protocol.

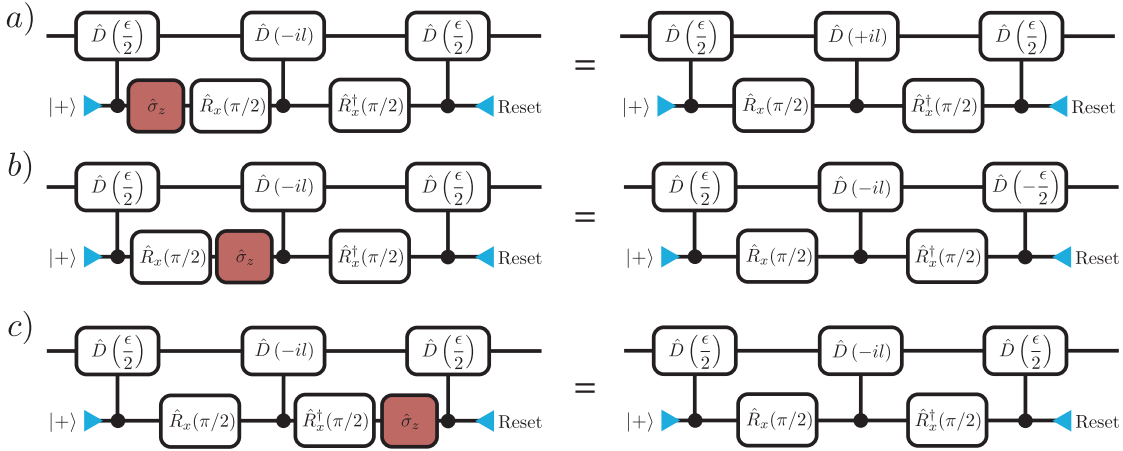

FIG. S13. Robustness of the protocol against dephasing errors. $\hat{\sigma}_z$ errors (red boxes) can occur during the first (a), second (b), or third (c) controlled displacements. An equivalent circuit for each of these cases is shown on the right. In all cases, the GKP logical information is preserved.

Dephasing errors during the first controlled displacement are equivalent to ancilla preparation errors, and we conclude that all protocols are robust to such state preparation errors. It is also interesting to draw a parallel between state preparation errors and the bath engineering picture. In the input-output model presented in Sect. IV C we considered a zero-temperature bath, but we note that ancilla preparation errors are equivalent to a finite-temperature bath [6]. Given a probability $p_{\rm e}$ of making a preparation error, the initial mixed state of the ancilla is given by

$$\rho_{\rm anc} = (1 - p_{\rm e})|+\rangle\langle+| + p_{\rm e}|-\rangle\langle-|. \tag{S107}$$

We can alternatively express that state as a thermal state [6]

$$\rho_{\rm anc} = \frac{1 + N}{2N + 1}|+\rangle\langle+| + \frac{N}{2N + 1}|-\rangle\langle-|, \tag{S108}$$

where the average number of bath excitations is set by the Bose-Einstein statistics $N = 1/(e^\beta - 1)$ with $\beta = T^{-1}$ the inverse (dimensionless) temperature. We can therefore compute an effective bath temperature from the preparation error probability,

$$\beta = \ln\left(\frac{1 - p_{\rm e}}{p_{\rm e}}\right). \tag{S109}$$

In order to protect the logical information contained in the GKP states, we should have $T = \beta^{-1} \ll 1$.

### C. Tunneling due to ancilla decoherence

As discussed above, both ancilla decay and dephasing can lead to large displacements errors. One side-effect of these large displacement errors is that they can compound to produce a displacement of size $\sim \pi/\epsilon$, leading to tunneling between sites of the superlattice. As a result, the confinement induced by the stabilization protocol is reduced in the presence of ancilla decoherence. For larger decay and dephasing rates, we correspondingly observe a slow increase in photon number during the stabilization.

## VIII. CONCATENATION OF GKP QUBIT WITH OTHER CODES

While the GKP stabilization scheme presented here leads to the encoding of highly coherent qubits, it does not provide a way to increase the lifetime of the logical information arbitrarily. Indeed, for a given error channel, Fig. 3 of the main Letter shows that there is an optimal GKP size that maximizes the fidelity

and increasing the size further leads to poorer performances, see the discussion in Sect. VII A 3. In order to reach the error levels required for long computations, we envision concatenation of the GKP encoding with some other qubit code.

A first option towards that goal is to combine the stabilization protocols with the Pauli measurements circuits presented in Sect. V B. In this case scenario, the GKP logical qubit is operated as a standard qubit, which allows the use of standard QEC techniques for the higher levels of the concatenated code. In other words, the bosonic nature of the GKP is not used in higher levels of the concatenation. In contrast, a second approach is to use the full, continuous information provided by measurements of the GKP stabilizers to improve the fidelity of the higher-level codes [11]. For example, concatenation with the surface code leads to the surface-GKP code which was studied in Refs. [12–14]. In both of these approaches, the stabilization protocol presented in this Letter can be used to enhance the lifetime of the base-level GKP qubit.

## IX. LATTICE RESHAPING

It is interesting to note that it is possible to adiabatically change the stabilization protocol in time such that the underlying lattice shape changes in time. For example, consider the change from square to hexagonal lattice,

$$
\begin{aligned}
L_\square &\to L_\bigcirc, \\
l_\square \begin{pmatrix} 0 & -1 \\ 1 & 0 \end{pmatrix} &\to l_\bigcirc \begin{pmatrix} 0 & -1 \\ \sin(\pi/3) & \cos(\pi/3) \end{pmatrix}.
\end{aligned}
\tag{S110}
$$

We parametrize an angle and a length as a function of time

$$
\begin{aligned}
\theta_t &= \frac{\pi}{2} - \frac{\pi}{6} \times \frac{t}{T}, \\
l_t &= \frac{l_\square}{\sqrt{\sin \theta_t}},
\end{aligned}
\tag{S111}
$$

where $t, T \in \mathbb{Z}$ with $T$ the total number of steps over which the lattice change is made. We then get a time-dependent lattice

$$
L_t = l_t \begin{pmatrix} 0 & -1 \\ \sin \theta_t & \cos \theta_t \end{pmatrix},
\tag{S112}
$$

that starts as a square lattice, $L_0 = L_\square$, and ends as an hexagonal lattice, $L_T = L_\bigcirc$. At all times, the lattice respects the correct commutation relation condition, $\det L_t = l_t^2 \sin \theta_t = 4\pi \, \forall \, t$. Fixing a GKP size $\Delta$, one can then use Fig. S6 to find the circuits corresponding to each $L_t$.

Consider a GKP state that is an exact $+1$ eigenstate of the finite-energy stabilizers at time $t$. This state can equivalently be thought of as a $+1$ eigenstate of the lattice at time $t + 1$ plus some displacement errors. If the difference between the lattice $L_t$ and $L_{t+1}$ is given by displacement errors much smaller than $\epsilon$, then the stabilization circuit at time $t + 1$ corrects these errors and the GKP state becomes a $+1$ eigenstate of the new stabilizers of $L_{t+1}$. By repeating this process and by choosing a parametrization such that $L_t$ and $L_{t+1}$ are close enough for all $t$, it is possible to reshape the GKP state to any desired lattice. Because the errors due to lattice changes are always correctable in this context, the logical information is preserved during the procedure. Note that for large GKP states, a small difference in lattice angle corresponds to large displacement differences at the edges of the lattice. Consequently, lattice reshapes require more steps as the GKP size increases.

---


    }

\end{document}